\documentclass[useAMS,usenatbib]{mn2e}
\usepackage[dvips]{graphicx}
\usepackage{amssymb,amsmath}
\usepackage{color}
\usepackage{url}

\title[Shape of Galaxies in SDSS/Galaxy Zoo]{The Intrinsic Shape of Galaxies in SDSS/Galaxy Zoo}
\author[S. Rodr\'iguez and N.D. Padilla]{Silvio Rodr\'iguez$^1$\thanks{E-mail: sirodrig@uc.cl} and Nelson D. Padilla$^{1,2}$\\
$^{1}$Departamento de Astronom\'\i a y Astrof\'\i sica, Pontificia Universidad Cat\'olica de Chile, Santiago, Chile\\
$^{2}$Centro de Astro-Ingenier\'ia, Pontificia Universidad Cat\'olica de Chile, Santiago, Chile}

\begin{document}

\date{Accepted --. Received --; in original form 2013 April 23}

\pagerange{\pageref{firstpage}--\pageref{lastpage}} \pubyear{2013}

\maketitle

\label{firstpage}

\begin{abstract}
  By modelling the axis ratio distribution of SDSS DR8 galaxies we find the intrinsic 3D shapes of spirals and ellipticals.  We use morphological information from the Galaxy Zoo project and assume a non-parametric distribution intrinsic of shapes, while taking into account dust extinction.
  
  We measure the dust extinction of the full sample of spiral galaxies and find a smaller value than previous estimations, with an edge-on extinction of $E_0 = 0.284^{+0.015}_{-0.026}$ in the SDSS $r$ band.  We also find that the distribution of minor to major axis ratio has a mean value of $ 0.267 \pm 0.009 $, slightly larger than previous estimates mainly due to the lower extinction used; the same affects the circularity of galactic discs, which are found to be less round in shape than in previous studies, with a mean ellipticity of $0.215 \pm 0.013$.

  For elliptical galaxies, we find that the minor to major axis ratio, with a mean value of $0.584 \pm 0.006$, is larger than previous estimations due to the removal of spiral interlopers present in samples with morphological information from photometric profiles.  These interlopers are removed when selecting ellipticals using Galaxy Zoo data.

  We find that the intrinsic shapes of galaxies and their dust extinction vary with absolute magnitude, colour and physical size. We find that bright elliptical galaxies are more spherical than faint ones, a trend that is also present with galaxy size, and that there is no dependence of elliptical galaxy shape with colour. For spiral galaxies we find that the reddest ones have higher dust extinction as expected, due to the fact that this reddening is mainly due to dust.  We also find that the thickness of discs increases with luminosity and size, and that brighter, smaller and redder galaxies have less round discs.

\end{abstract}


\section{Introduction}

The study of the real shape of galaxies started as early as galaxies were first classified into morphological types with Hubble \citep{hubble2}, who measured the $b/a$ ratios for galaxies classified as ellipticals. Many quantitative studies were performed before the invention of the CCD cameras.  For instance \cite{sandage}, using the projected axial ratios from 254 spiral galaxies from the Catalogue of Bright Galaxies \citep{deVacouleurs}, determined that the disc of spirals were circular, with a disc thickness defined as $\gamma \equiv C/A$, with $A$,$B$ and $C$ the major, medium and minor axis, respectively, given by $\gamma=0.25$. A step further was given by \cite{binggeli}, \cite{benacchio} and \cite{binney}, who did not assume that the disc of spiral galaxies was circular, and measured their ellipticity, defined as $\epsilon \equiv 1-B/A$. They found that the mean value of $\epsilon$ was given by $\epsilon=0.1$.

\cite{lambas} found that the distribution of $\epsilon$ is well fit by a one-sided Gaussian distribution centred on $\epsilon=0$. They used a sample of $\sim$ 13 000 galaxies from the APM galaxy survey \citep{apm} to find a dispersion of $\sigma_{\epsilon}=0.13$ and a mean of $\epsilon$ given by $\langle\epsilon\rangle=0.1$. \cite{rix} studied a sample of face-on spirals (selected kinematically) in detail, and they found an ellipticity of $\epsilon=0.045$.

Other studies take into count the fact that from spatially resolved observations of the inner kinematics it is possible obtain the 3D shape of a galaxy \citep{binney0,franx,statlera,statlerb,statler2,statler3,statler4}. For instance, \cite{andersen1} and \cite{andersen2} used this method in 24 face-on spirals to obtain a mean ellipticity of $\langle \epsilon \rangle = 0.076$, similar to the value obtained by \cite{rix}. However, the fact that in both cases the sample is composed entirely of face-on objects may introduce systematic biases.

The advent of the Sloan Digital Sky Survey \citep[SDSS,][]{york} allowed studies with larger number of galaxies with high-quality photometry. \cite{ryden1}, using a sample of spirals from SDSS Data Release 1 \citep[DR1,][]{dr1} chosen to minimize systematics due to seeing, found that the distribution of galactic disc ellipticities can be well fitted by a log-normal distribution with a mean of $\langle\log\epsilon\rangle = -1.85$ and a standard deviation of $\sigma_{\log\epsilon}=0.89$. \cite{ryden2} go even further. Using data from SDSS Data Release 3 \citep[DR3,][]{dr3}, they fit the distribution of axis ratios of both ellipticals and spirals to triaxial models. Assuming a uniform triaxiality (i.e. all galaxies are either prolate, triaxial or oblate), they found that spirals and ellipticals are consistent with oblate spheroids and that high-luminosity ellipticals tend to be rounder than low-luminosity ellipticals.

Elliptical galaxies were once believed to be axisymmetric oblate spheroids, until \cite{bertola} discovered that their rotation velocities were insufficient to support such a geometry. \cite{binney-1} suggested that ellipticals could be well described by a triaxial ellipsoid but \cite{davies} found that small ellipticals are better fitted by oblate spheroids. This variety of shapes made it difficult to obtain their intrinsic shapes using only their apparent images, an approach which was often used for large samples of ellipticals. In such cases it is necessary to assume triaxiality as in \cite{ryden2}, or, following \cite{binney2}, to use the misalignment between the internal isophotes of individual elliptical galaxies. \cite{mendez} extended the study of the intrinsic shapes of spheroids to bulges in spiral galaxies. They found that the bulge shape is consistent with a mean axial ratio in the equatorial plane of $\langle B/A \rangle = 0.85$.

The distribution of the shapes of spiral galaxies will be additionally affected by dust extinction \citep{holmberg}. Optically thick dust obscuration aligned in the rotational plane of spirals will cause edge-on objects to appear systematically fainter. This effect will introduce a bias in flux-limited samples.  It is important to quantify this effect to understand the true luminosities of galaxies, the distribution of the interstellar medium (ISM) and the relationship between optical and infrared emission of galaxies \citep[for reviews, see][]{davies2,calzetti}.

Studies of the brightness of galaxies as a function of axial ratio should allow the effects of dust to be quantified. \cite{valentijn}, studying the shapes and brightness of 16 000 galaxies from digitalized photographic plates, found indication of an optically thick component in disc galaxies, extending beyond the apparent optical extent of the galaxy. However, \cite{burnstein} and \cite{choloniewski} showed that Valentijn's results were due in part to selection effects, and found that the diameters of galaxies were independent of the inclination \citep[for further discussion, see][]{davies3,valentijn2}. Expanding on the effects of selection biases with the inclination \cite{peletier} emphasized that dust opacity may depend on galaxy luminosity. Following this \cite{tully} found a difference between face-on and edge-on luminous galaxies of 1.3 mag in the $R$ band, but found no important difference in faint galaxies. \cite{holwerdaa,holwerdab} measure the number of galaxies seen trough galactic discs using images from the \textit{Hubble Space Telescope} WFPC2 archival data to measure the opacities of spiral discs. This method has been previously applied to ground-based data in other works \citep[for example][]{zaritski,nelson,keel}. Using a different method, \cite{valotto} use the inner part of the rotation curves for spiral galaxies to derive dust extinction.

Following other methods, a number of groups have studies the variation of the galaxy properties with the inclination angle with respect to the line of sight, or directly with projected galaxy shapes, to obtain the dust extinction in spirals galaxies. \cite{shao-etal}, using spiral galaxies from the SDSS DR2 \citep{dr2}, measure the dust extinction by studying the luminosity function (LF) of galaxies with different inclination angles, and using the intrinsic galaxy shape obtained from the analysis of the distribution of projected axis ratios. They claim that the decrease in LF with the increasing the inclination is an effect of the dust extinction, where the disc optical depth is roughly proportional to the cosine of the inclination angle. But, in their calculation, they do not take into account the effect of the dust in the distribution of the projected axis ratios. Using a sub-sample of $\sim$ 78 000 galaxies from SDSS DR6 \citep{dr6}, \cite{ryden3} find similar results for the dependence of extinction on the projected shape. They use these results to define an unbiased sample of spiral galaxies, and obtain the intrinsic shapes of this sample. They found that these galaxies are consistent with flattened disc as was found by previous works, but the definition of the sample makes it difficult to compare these results with previous estimates. \cite{maller}, using the NYU-VAGC \citep{nyu-vagc} catalogue, which combines data from SDSS and the Two Micron All Sky Survey \citep{2MASS}, study the variation of galaxy properties with inclination. They found a median extinction over the whole sample of 0.3 mag in the $g$ band. \cite{driver} study the dependence of the LF with inclination, by decomposing their sample of Galaxies in the Millennium Galaxy Catalogue \citep{MGG1,MGG2} into bulge and disc components, and are able to obtain the residual face-on attenuation.

In elliptical galaxies, dust has also been found. For example, \cite{ebneter} used colour maps to find evidence of dust in more than 30 \% of their sample of ellipticals, and 2.5 \% of the galaxies showed evidence of a dusty disc. But the amount of dust in ellipticals is smaller than the amount of dust in spirals. \cite{knapp} found that the amount of dust in an elliptical galaxy is between 1 - 10 \% of the dust in a spiral of similar luminosity \citep[see also][]{goudfrooij,krause,leeuw}. \cite{temi}, using far-infrared observations, placed constraints in the mass of dust in ellipticals in the range $M_{dust} = 10^5-10^7 M_{\odot}$ h$^{-1}$ (h is the Hubble constant in units of 100 km s$^{-1}$ Mpc$^{-1}$). This mass represents $\sim 10^{-6}$ of the stellar mass. In spirals, the fraction of dust mass over stellar mass is of the order of $5\times10^{-3}$ \citep{stevens}.

\cite{nym} (hereafter PS08) took in count the effect of dust obscuration in the study of the intrinsic shape of galaxies. They used a model with a normal distribution for $\gamma$ and a log-normal distribution for $\epsilon$, and use non-uniform distributions of inclinations for their samples to take into account the effect of dust, parametrized with the edge-on extinction $E_0$, which removes edge-on galaxies from their samples. For elliptical galaxies, they set the dust extinction to 0. They fit the dust and shape parameters using the observed $b/a$ distribution of SDSS DR6 \citep{dr6} galaxies. To separate the sample into spirals and ellipticals, PS08 use the $fracDeV$ parameter, defined in \cite{dr3}. This parameter indicates whether the luminosity profile is closer to exponential (lower $fracDeV$) or de Vaucouleurs (higher $fracDeV$). Using this model PS08 obtain a mean for $\gamma$ given by $\langle\gamma\rangle=0.21\pm0.02$, with dispersion $\sigma_{\gamma}=0.05\pm0.015$, a mean for $\epsilon$ given by $\langle\ln\epsilon\rangle=-2.33 \pm 0.13$, with dispersion $\sigma_{\ln \epsilon}=0.78\pm0.16$, and $E_0=0.44\pm0.24$ for spirals. For ellipticals they obtain $\langle\gamma\rangle=0.43\pm0.06$, $\sigma_{\gamma}=0.21\pm0.02$, $\langle\ln\epsilon\rangle=-2.2\pm0.1$ and $\sigma_{\ln \epsilon}=1.4 \pm 0.1$. These results are consistent with oblate spheroids.

Following PS08, \cite{ssg} (SSG10) made a similar analysis for AGN host galaxies, but without considering dust. They found that the distributions of $\sigma$ and $\epsilon$ for spirals fit well a sum of gaussians distribution, instead a single Gaussian or log-Gaussian distribution.

\cite{lagos} apply the same analysis as PS08 to a sample of AGN host galaxies from the SDSS DR7 \citep{dr7}. They found that type I and type II AGN have a similar intrinsic shape, consistent with the unified AGN model, with a $\gamma=0.23\pm0.08$ for spirals and $\gamma=0.6\pm0.24$ for ellipticals. Also, they found that type I AGN tend to be face-on, while type II AGN tend to be edge-on, albeit with lower statistical significance.

This work builds upon analysis made in PS08, with the aim to determine the real 3D shape of galaxies with different intrinsic properties using data available in the Sloan Digital Sky Survey (SDSS), using an improved non-parametric distribution of $\gamma$ and $\epsilon$.

With respect to PS08, this work presents three major changes,

\begin{itemize}
  \item We separate the sample of spiral and elliptical galaxies using data obtained by The Galaxy Zoo project \citep{lintott} in addition to the use of the $fracDeV$ parameter.
  \item We calculate the value of the dust extinction ($E_0$) using the luminosity function, in a method unrelated to the fit of distributions of $b/a$.
  \item We use a linear combination of 10 gaussian distributions, each one with a fixed mean and dispersion, to obtain the total distribution of $\gamma$ following SSG10,  and modify the percentage of galaxies belonging to each gaussian, instead of using a single gaussian distribution for $\gamma$. Which can be considered to be to a non-parametric distribution. For $\epsilon$ we use 10 gaussians or a log-normal distribution depending on the sample.
\end{itemize}

The model for dust extinction is the same model used in PS08. Throughout this work we assume a standard $\Lambda$CDM cosmology, with matter density parameter $\Omega_m=0.3$ and cosmological constant $\Omega_{\Lambda}=0.7$.

This work organized as follows: Section \ref{sec:sample} describes the sample and different sub-samples of galaxies used in this work. In Section \ref{sec:model} we describe the main ideas of the model used to find the intrinsic shapes and the extinction. Section \ref{sec:result1} presents the results of the main sample of galaxies, and these results are compared with PS08, Section \ref{sec:result2} shows the results for different sub-samples separated by intrinsic properties of the galaxies. Section \ref{sec:conclusions} summarises and discusses our results.

\section{The Data}
\label{sec:sample}
The Sloan Digital Sky Survey (SDSS) is an imaging survey which covers approximately 14555 deg$^2$ of the sky, with spectra and five-band photometry of a large number of objects, including more than 860 000 galaxy spectra in the Data Release 8 \citep[DR8,][]{dr8}. The description of the technical details is in \cite{york}.

Galaxy Zoo is a project that aims to classify the morphology of a large number of SDSS galaxies. To reach this goal, they initially used galaxies from SDSS Data Release 6 \citep[DR6,][]{dr6}, and showed three-colour images of these galaxies to internet volunteers, who did the classification by eye. The description of the project can be found in \cite{lintott}, and the first data release from the Galaxy Zoo project is available in the SDSS web site along with data for the SDSS DR8, and is described in \cite{lintott2}.

In this work we use data for all the galaxies from SDSS DR8 with both spectroscopic and Galaxy Zoo information available. We K-correct the galaxy magnitudes using V3.2 of the code described in \cite{kcorrect}. The following subsections describe the cuts applied to these objects to remove biases and separate sub-samples.

\subsection{Removing systematic biases in the sample}
\begin{figure}
  \begin{center}
    \includegraphics[width=.44\textwidth]{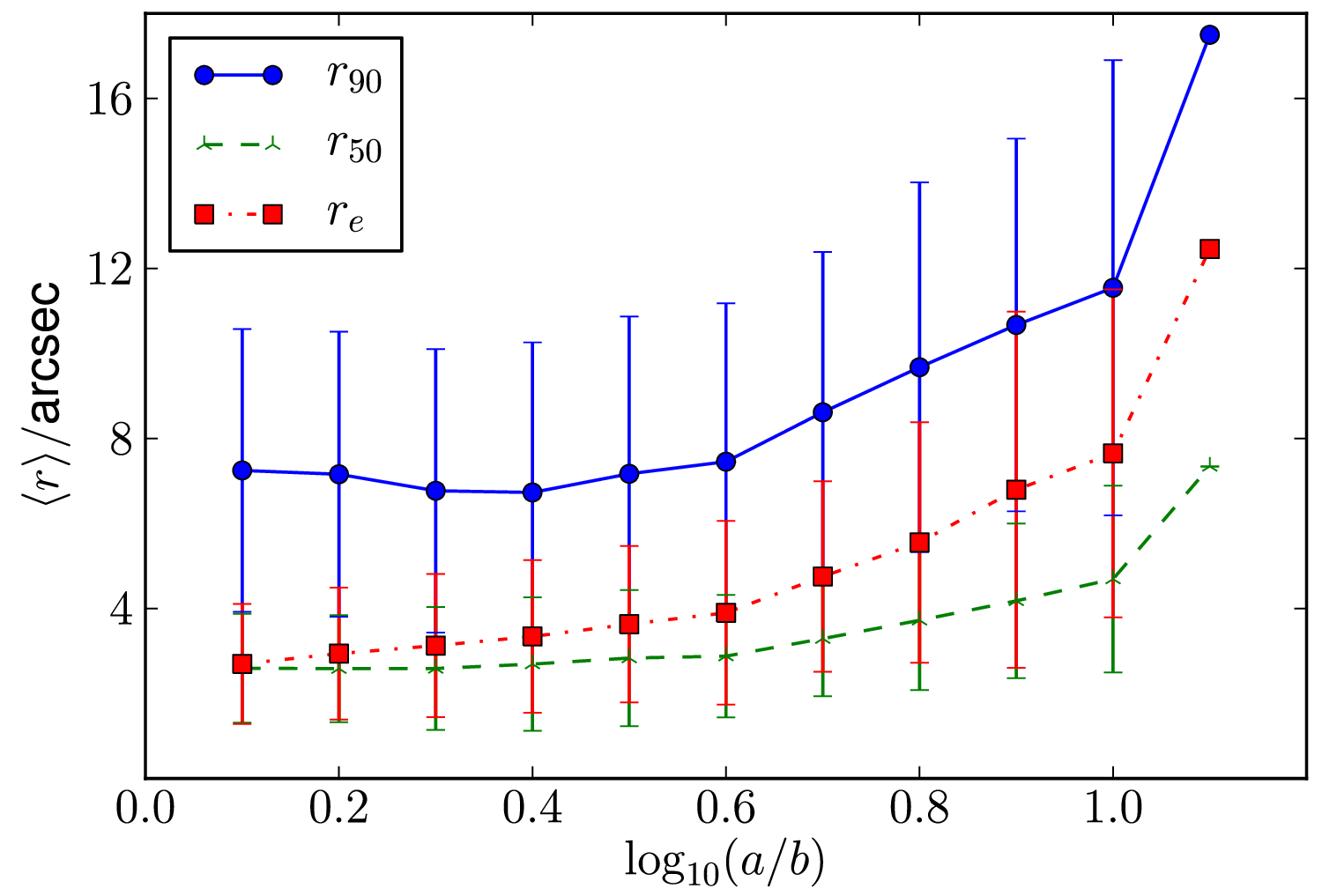}
    \caption{Mean value of photometric radius for different ranges of $b/a$. We show results using $r_{90}$, $r_{50}$ and $r_e$ as shown in the key.\label{fig:r90_re}}
  \end{center}
\end{figure}

\cite{masters} analyse Galaxy Zoo galaxies and show that there is a relationship between $r_{90}$ and the $b/a$ measured directly from imaging due to the effect of seeing. Using the SDSS data, we calculate the value of $\langle r_{90} \rangle$, for different $b/a$ ranges. Figure \ref{fig:r90_re} shows the results.
As can be seen, we find a relationship between $r_{90}$ and $b/a$, in which the galaxies with $\log(a/b)>0.8$ tend to have a larger $r_{90}$.  However, $r_{90}$ can be a noisy size estimator, and therefore we repeat
the analysis with $r_{50}$; this size estimator shows the same trend.  Since both PS08 and this work use model $b/a$ ratios calculated using an exponential or de Vaucouleurs models convolved with the seeing,
we repeat the analysis using $r_e$ (apparent radius obtained fitting an exponential model).  The trend found with $r_{90}$ is still present with $r_e$.   Therefore, from this point on we do not take in count galaxies with $\log(a/b)>0.8$, that is $b/a<0.15$, to avoid this systematic effect. The removed galaxies represent less than 0.05 \% of the sample.

\begin{figure}
  \begin{center}
    \includegraphics[width=.44\textwidth]{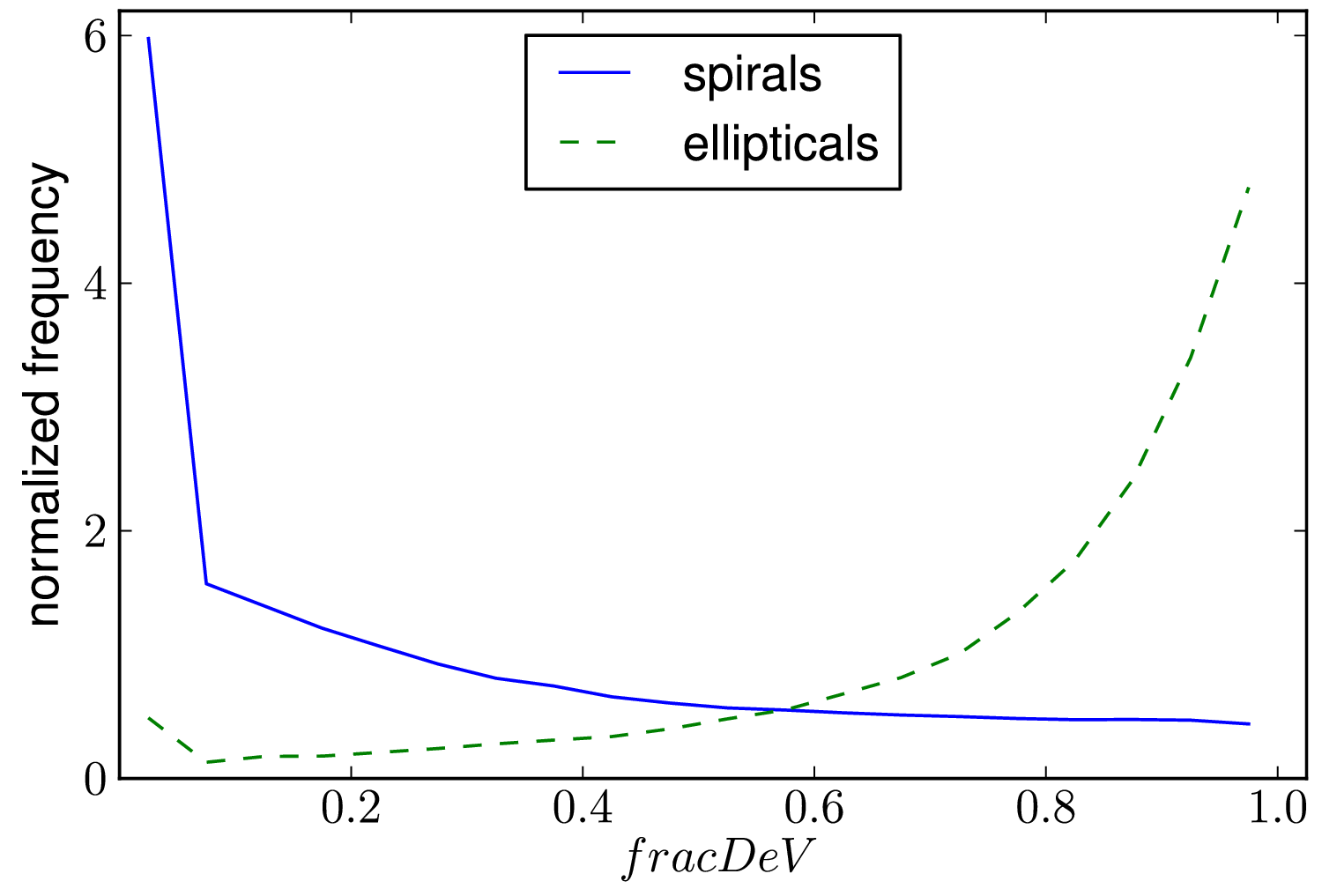}
    \caption{Distribution of $fracDeV$ parameter for the spiral and elliptical samples when the \textit{greater} criterion is used.\label{fig:fdeV-hist}}
  \end{center}
\end{figure}

Following \cite{lintott2}, we decided to use the sample of spirals and ellipticals classified using the \textit{greater} criterion, that is, if a galaxy has been classified as spiral by more than 50\% of the volunteers, then the galaxy is considered a spiral, and the same for ellipticals. But Galaxy Zoo classification is not enough, due to the fact that galaxies which are at large distances will appear as a blur. As a consequence galaxies with an exponential luminosity profile could be classified as ellipticals and galaxies with a de Vaucouleurs luminosity profile could be classified as spirals. Figure \ref{fig:fdeV-hist} shows the  $fracDeV$ distribution for the sample when the \textit{greater} criterion is applied, for spirals and ellipticals; the figure shows that there is a number of galaxies in the spiral sample with a de Vaucouleurs luminosity profile, and galaxies in the elliptical sample with a exponential luminosity profile. To avoid this, we use the $fracDeV$ parameter (in SDSS $r$ band) as a second filter, so a galaxy which is in the spiral sample according the \textit{greater} criterion and has a $fracDeV < 0.8$ (the limit used in PS08) will be considered a spiral galaxy, and a galaxy which is in the elliptical sample according the \textit{greater} criterion and has a $fracDeV > 0.8$ will be considered an elliptical galaxy.

\begin{figure}
  \begin{center}
    \includegraphics[width=.44\textwidth]{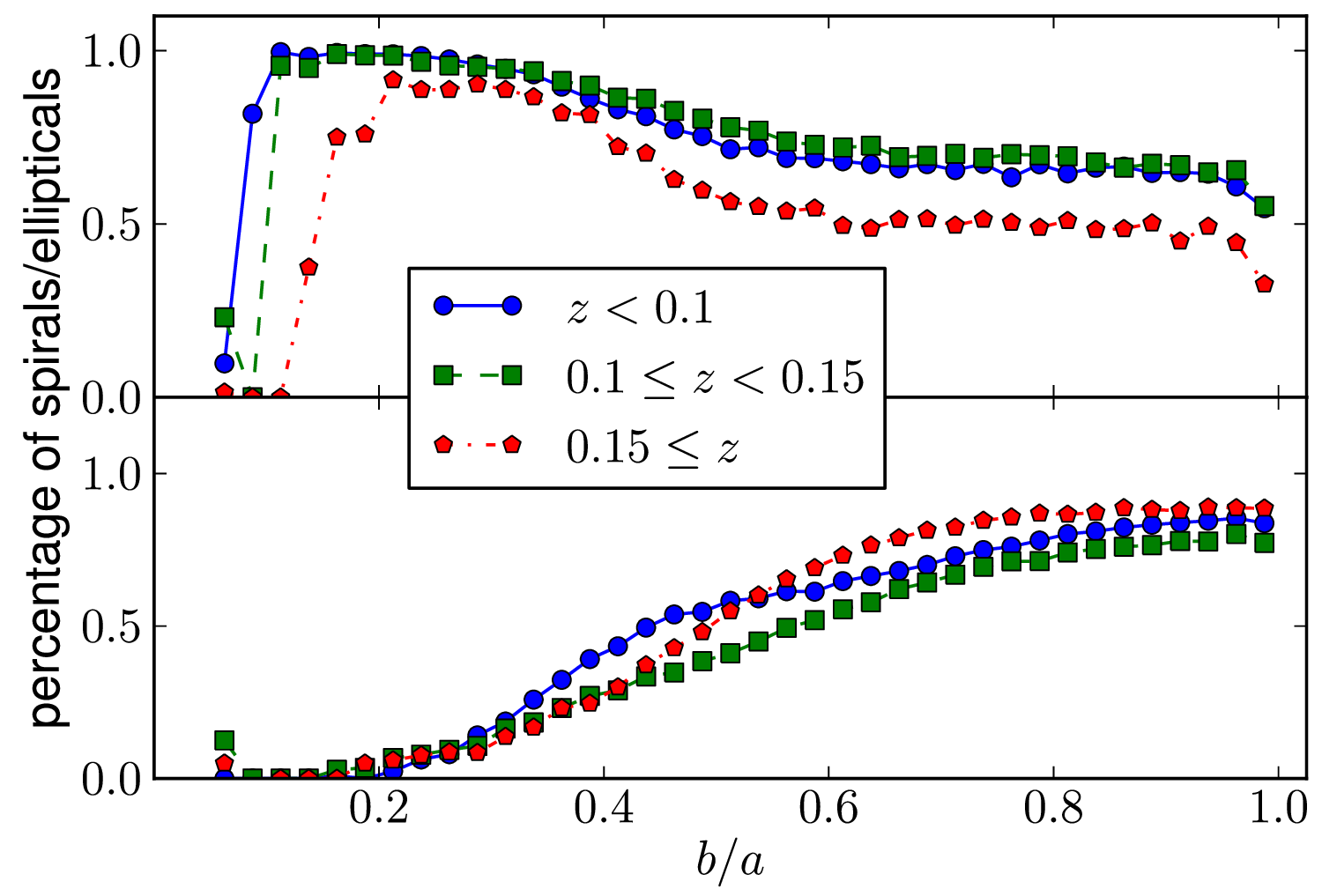}
    \caption{Top: Percentage of galaxies considered as spirals by Galaxy Zoo in the sample of galaxies with $fracDeV < 0.8$, in bins of $b/a$ for different redshift slices. Bottom: same as the top panel for ellipticals in the sample of galaxies with $fracDeV > 0.8$.\label{fig:p_byz}}
  \end{center}
\end{figure}

The use of $fracDeV$ as a second morphology estimator is not enough to clean the sample of the bias due the classification by eye at large $z$. Figure \ref{fig:p_byz} shows the percentage of galaxies with an exponential profile considered as spirals by the volunteers of Galaxy Zoo in bins of $b/a$, at different redshifts. And the same for ellipticals in the galaxies with a de Vaucouleurs profile. It is clear that the percentage of spirals is smaller in galaxies at $z \geq 0.15$, in particular the percentage of face-on spirals. Ellipticals are not affected by this effect.  We choose to remove from the sample all the galaxies (spirals and ellipticals for consistency) with $z \geq 0.15$.

In addition, the sample has a cut-off in absolute magnitude, with a magnitude limit in the r-band $M_r < - 19.77$, that is, the minimum magnitude for a galaxy to be included in the SDSS survey at $z=0.1$; this cut allows the inclusion of brighter galaxies at higher $z$. With these considerations, the spirals sample contains 92923 galaxies, and the ellipticals sample contains 112100 galaxies.

\begin{figure}
  \begin{center}
    \includegraphics[width=.44\textwidth]{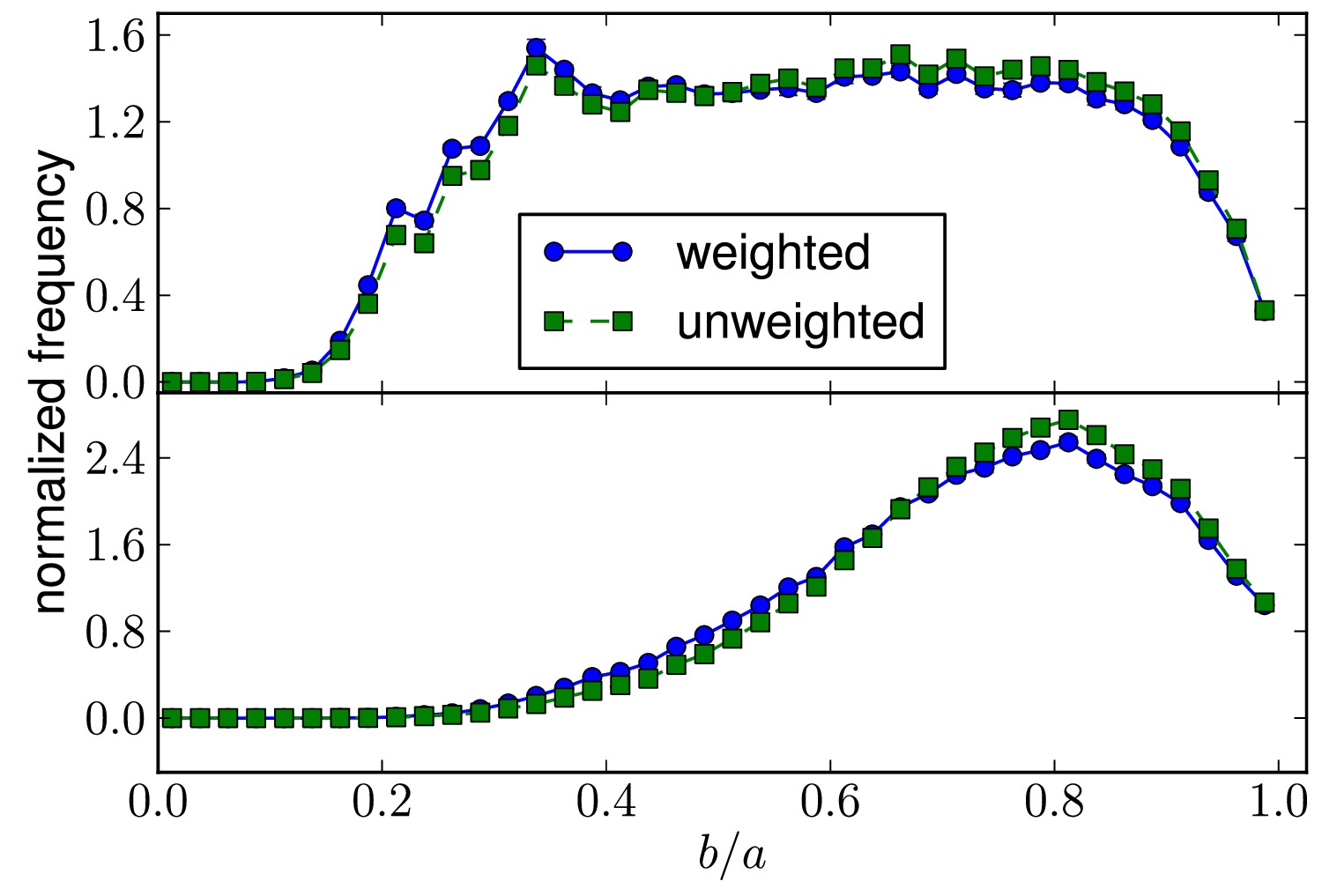}
    \caption{Top: distribution of $b/a$ for spirals. Bottom: distribution of $b/a$ for ellipticals.  In both panels the connected circles show the distributions weighted by $1/V_{max}$ to account for the flux limit of the sample.  The squares show the unweighted distributions. \label{fig:badist}}
  \end{center}
\end{figure}

Figure \ref{fig:badist} shows the $b/a$ distribution for these samples weighted by $1/V_{max}$ , where $V_{max}$ is the maximum volume corresponding to the distance at which a galaxy with a given absolute magnitude enters the flux-limited catalogue ($m_r\leq $17.77), and the distribution of $b/a$ without weighting. In further analysis, the $1/V_{max}$ weight will be used always.

An important part of this work is not only to measure the intrinsic shapes and the dust of the total the sample of spirals and ellipticals, it is also important to verify how the shape and the dust obscuration change with the intrinsic properties of the galaxies. For this reason we separate the spiral and elliptical samples into sub-samples, these sub-samples will be described next.

\subsubsection{Spirals}

\begin{figure*}
  \begin{center}
    \includegraphics[width=.85\textwidth]{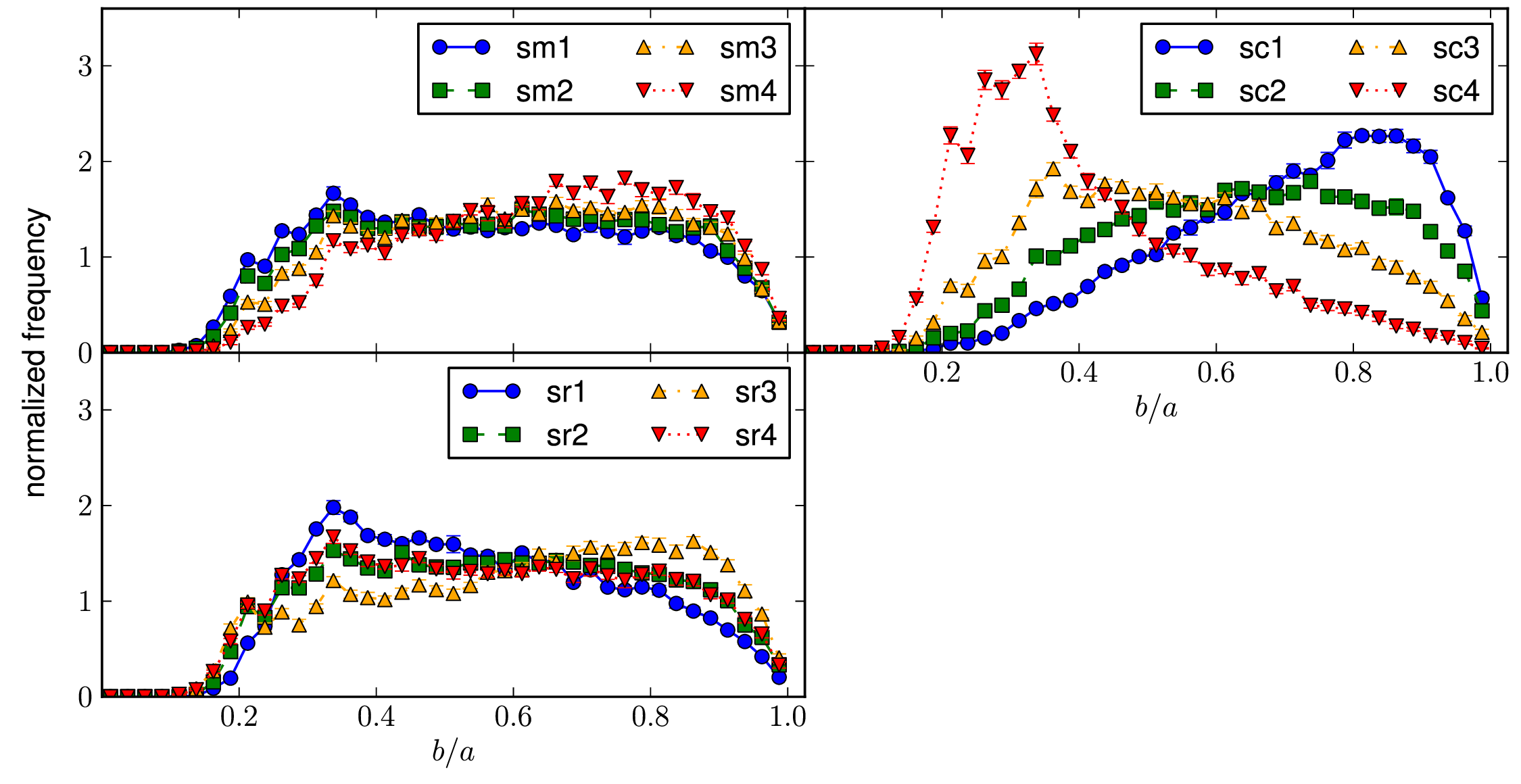}
    \caption{Distribution of $b/a$ for the spiral sample. Top left: Separated by $M_r$. Top right: Separated by $g-r$. Bottom left: Separated by $R_{50}$.\label{fig:ba-fordist}}
  \end{center}
\end{figure*}

Figure \ref{fig:ba-fordist} shows the $b/a$ distribution for the spiral sample, separated by absolute magnitude $M_r$, $g-r$ colour and radius $R_{50}$. Table \ref{table:not_exp} shows the naming convention for the sub-samples that will be used in the rest of this paper. The boundaries in the sub-samples were initially chosen to produce quartiles.  However, we apply cuts in redshift afterwards which changes the number of galaxies per sub-sample, which are listed in Table \ref{table:not_exp}.

\begin{table}
  \begin{center}
    \caption{Notation for the sub-samples of spirals rand number of galaxies per sub-sample.\label{table:not_exp}}
    \begin{tabular}{ccc}
      \hline
      Sample & Limits & Number\\\hline
      sm1 & $-20.14<M_r-5\log_{10} \textmd{h}$        & 26517 \\
      sm2 & $-20.5<M_r-5\log_{10} \textmd{h} <-20.14$ & 26493 \\
      sm3 & $-20.91<M_r-5\log_{10} \textmd{h}< -20.5$ & 25255 \\
      sm4 & $M_r-5\log_{10} \textmd{h} <-20.91$       & 14655 \\\hline
      sc1 & $g-r<0.58$                                & 23513 \\
      sc2 & $0.58<g-r<0.68$                           & 22915 \\
      sc3 & $0.68<g-r<0.79$                           & 22828 \\
      sc4 & $0.79<g-r$                                & 23666 \\\hline
      sr1 & $ R_{50}/\textmd{kpc h} <3.2$             & 26294 \\
      sr2 & $3.2<R_{50}/\textmd{kpc h}<3.88$          & 25628 \\
      sr3 & $3.88<R_{50}/\textmd{kpc h}<4.74$         & 23757 \\
      sr4 & $4.74<R_{50}/\textmd{kpc h}$              & 17244 \\\hline
    \end{tabular}
  \end{center}
\end{table}

Notice that the sub-samples were constructed by applying cuts on only one galaxy property at a time, and therefore the other two properties could show variations among the sub-samples (of increasing luminosity, for instance).  We will investigate this further in Section \ref{ssec:interdependence}, where we will attempt to make samples with variations in only one galaxy property. 

\begin{figure}
  \begin{center}
    \includegraphics[width=.44\textwidth]{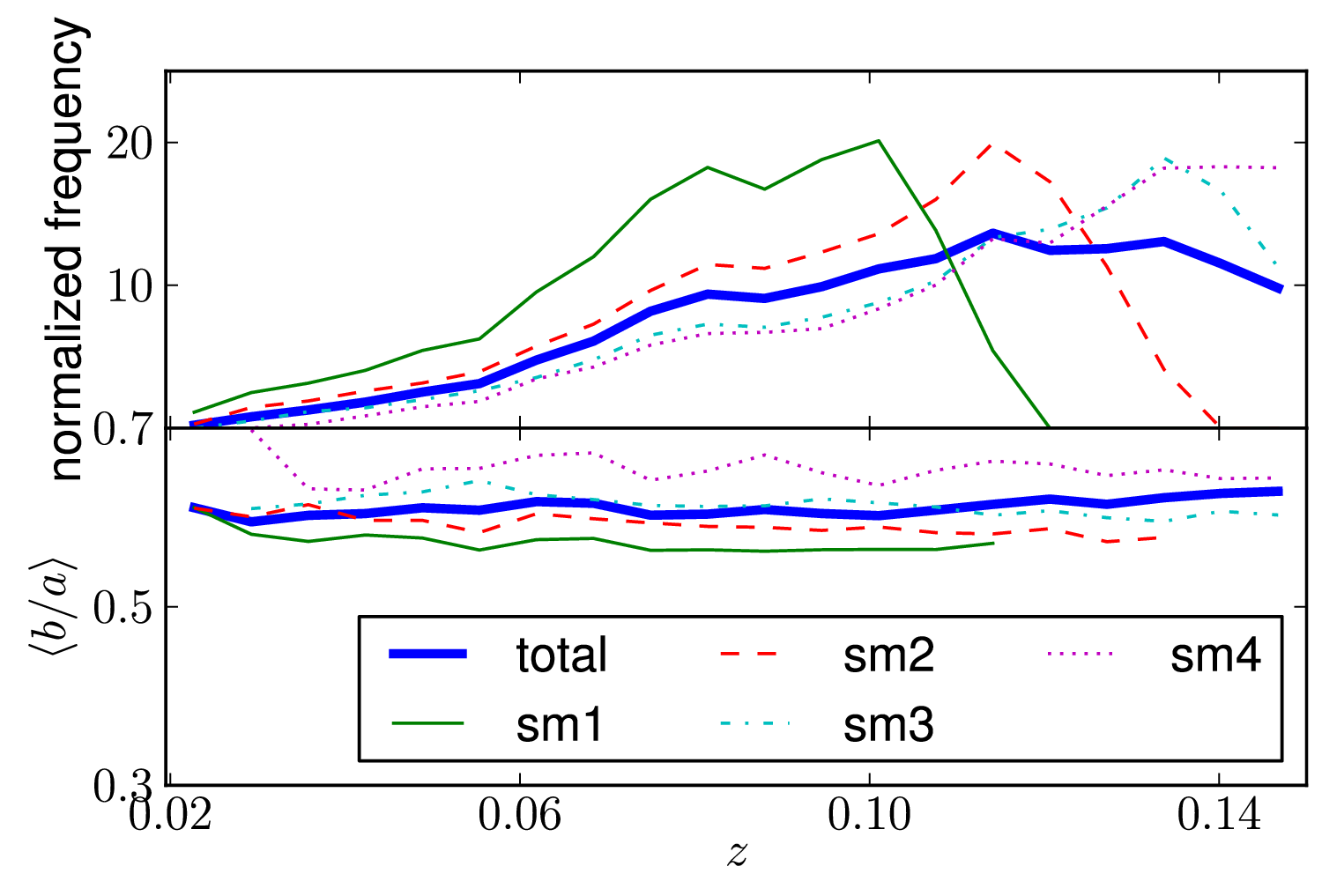}
    \caption{Top: Distribution of redshifts for the total sample of spirals, and for different sub-samples separated by $M_r$. Bottom: Relationship between redshift and $\langle b/a \rangle$ for the total sample of spirals, and for different sub-samples separated by $M_r$.\label{fig:exp-z}}
  \end{center}
\end{figure}

As the main morphological classification was done by human eyes (those of Galaxy Zoo volunteers), it is natural to check the sample for biases.  For instance, it is possible that we may only have spiral galaxies over a given magnitude range. The top panel in Figure \ref{fig:exp-z} shows the distribution of $z$, with the total distribution and the distribution of different sub-samples of spirals separated by $M_r$. This distribution is not different to the upper panels in Figure 1 of PS08 (within the range of $z$).
The bottom panel in Figure \ref{fig:exp-z} shows the relationship between $z$ and $\langle b/a \rangle$ for the total sample of spirals and for different sub-samples separated by magnitude. We can see that $\langle b/a \rangle$ does not change with $z$ apparently ruling out any further biases. 
Therefore we do not place further cuts on these samples.

\subsubsection{Ellipticals}
\begin{figure*}
  \begin{center}
    \includegraphics[width=.85\textwidth]{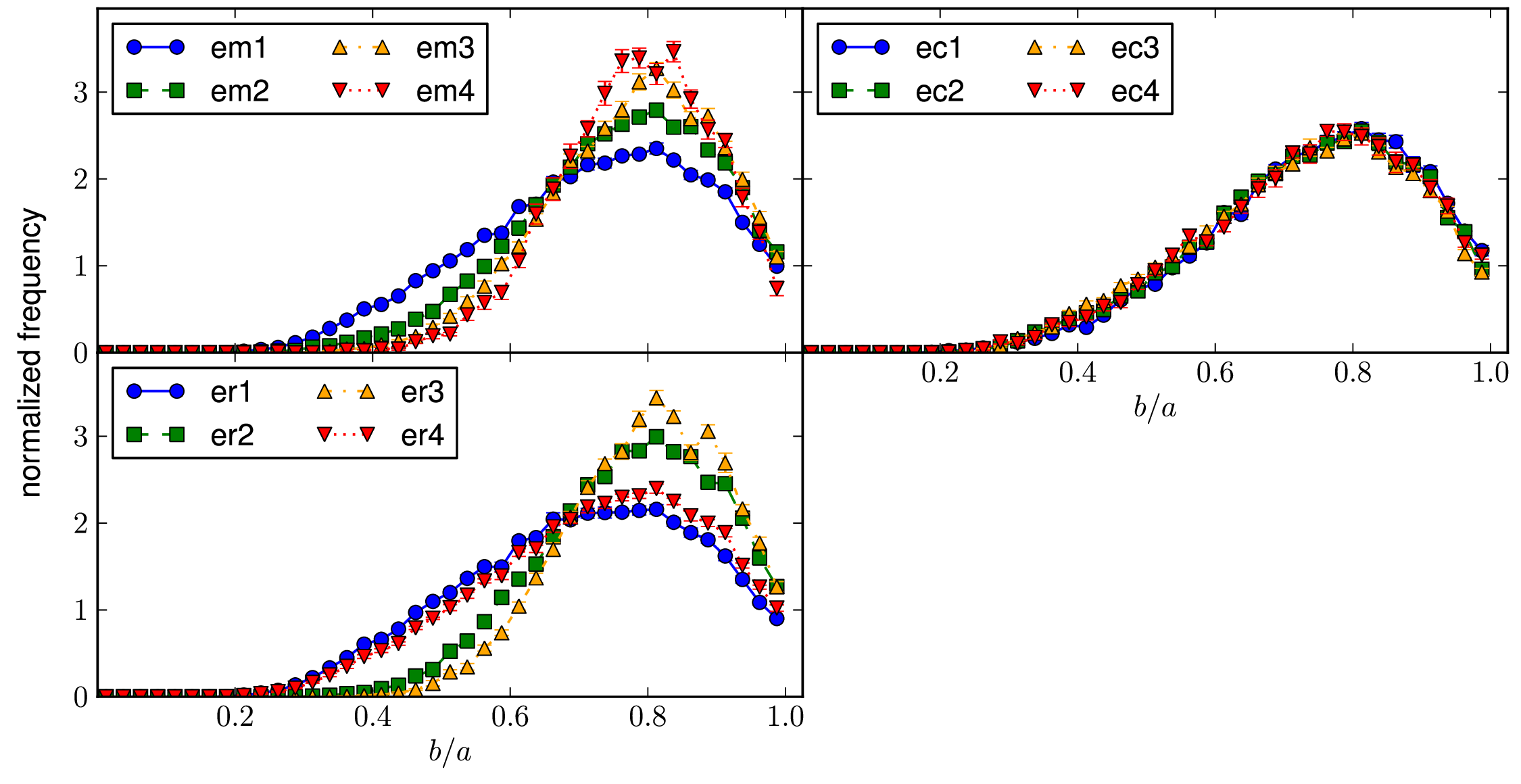}
    \caption{Distribution of $b/a$ for the elliptical sample. Top left: Separated by $M_r$. Top right: Separated by $g-r$. Bottom left: Separated by $R_{50}$.\label{fig:ba-fordist2}}
  \end{center}
\end{figure*}

Figure \ref{fig:ba-fordist2} shows the $b/a$ distribution for the elliptical sample separated by different galaxy properties ($M_r$,$g-r$ and $R_{50}$). Table \ref{table:not_deV} presents the naming convention for the sub-samples. As in the case of spiral galaxies, the sub-samples are quartiles of the total sample with subsequent cuts in redshifts.  The number of galaxies in each sub-sample is listed in Table \ref{table:not_deV}.

\begin{table}
  \begin{center}
    \caption{Names of sub-samples of ellipticals and number of galaxies per sub-sample.\label{table:not_deV}}
    \begin{tabular}{ccc}
      \hline
      Sample & Limits & Number \\\hline
      em1 & $-20.57<M_r-5\log_{10} \textmd{h}$        & 45246 \\
      em2 & $-21.12<M_r-5\log_{10} \textmd{h}<-20.57$ & 39454 \\
      em3 & $-21.6<M_r -5\log_{10} \textmd{h}<-21.12$ & 20110 \\
      em4 & $M_r -5\log_{10} \textmd{h} <-21.6$       & 7289 \\\hline
      ec1 & $g-r<0.93$                                & 30354 \\
      ec2 & $0.93<g-r<0.97$                           & 31836 \\
      ec3 & $0.93<g-r<1.01$                           & 29673 \\
      ec4 & $1.01<g-r$                                & 20237 \\\hline
      er1 & $R_{50}/\textmd{kpc h}<2.36$              & 44420 \\
      er2 & $2.36<R_{50}/\textmd{kpc h}<3.28$         & 37442 \\
      er3 & $3.28<R_{50}/\textmd{kpc h}<4.51$         & 21448 \\
      er4 & $4.51<R_{50}/\textmd{kpc h}$              & 8790  \\\hline
    \end{tabular}
  \end{center}
\end{table}

The redshift distribution and the relation between redshift and $\langle b/a\rangle$ for ellipticals, including the total sample and the sub-samples separated by $M_r$ are qualitatively similar to the result for the spiral sample. The elliptical sample is also unbiased in this respect.

\subsection{Comparison Samples}
\begin{figure}
  \begin{center}
    \includegraphics[width=.44\textwidth]{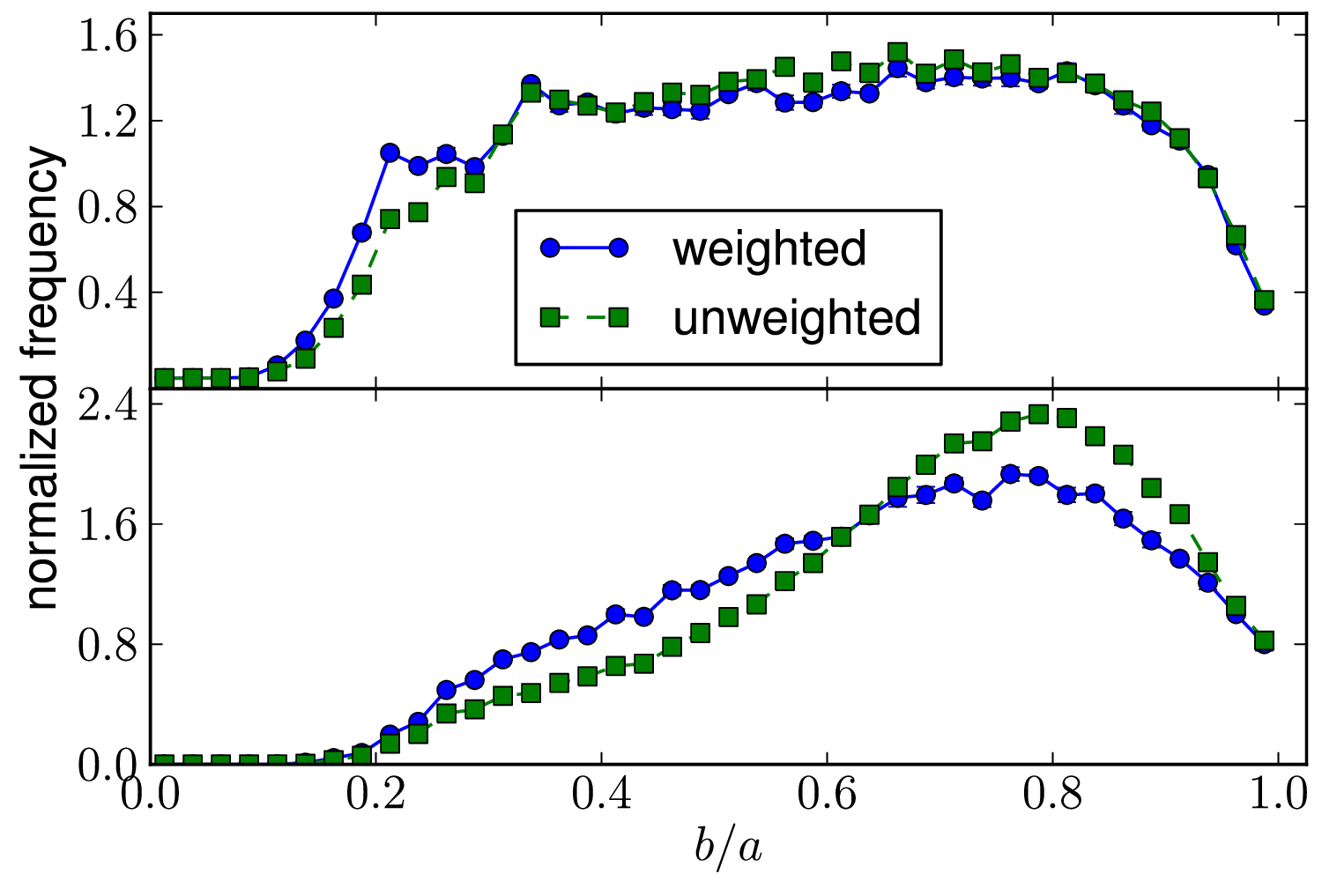}
    \caption{Distribution of $b/a$ for the comparison sample. Top: Distribution of $b/a$ for spirals. Bottom: Distribution of $b/a$ for ellipticals.  Symbols are as in Figure \ref{fig:badist}\label{fig:badist_comp}}
  \end{center}
\end{figure}

In order to quantify the impact of Galaxy Zoo morphologies, we use samples of galaxies separated between spirals and ellipticals only using the $fracDeV$ parameter obtained from SDSS DR6. As this sample is not affected by the bias introduced by the Galaxy Zoo, we do not impose limits on $z$. These samples are the same as those in PS08. Figure \ref{fig:badist_comp} shows the distribution of $b/a$ for these samples.

\section{Model parameters and best fit search}
\label{sec:model}
Following PS08, the model uses the distributions of $\gamma$ and $\epsilon$ to determine the distribution of $b/a$ for Monte-Carlo galaxies. The relationship between $\gamma$ and $\epsilon$ with $b/a$ is given by \citep{binney}
\begin{equation}
  \frac{b}{a} = \left[ \frac{V+Z-\sqrt{(V-Z)^2+W}}{V+Z+\sqrt{(V-Z)^2+W}} \right] \label{eq:ba}
\end{equation}
Where $V$, $W$ and $Z$ are given by
\begin{equation}
  V=[1-\epsilon(2-\epsilon)\sin^2\varphi]cos^2\theta+\gamma^2\sin^2\theta \label{eq:V}
\end{equation}
\begin{equation}
  W=4\epsilon^2(2-\epsilon)^2\cos^2 \theta \sin^2\varphi \cos^2\varphi \label{eq:W}
\end{equation}
\begin{equation}
  Z=1-\epsilon(2-\epsilon)\cos^2\varphi \label{eq:Z}
\end{equation}

Here, $\varphi$ and $\theta$ are the angles that characterize the orientation of the galaxy relative to the line of sight. $\varphi$ will be treated as a random value between 0 and $2\pi$, and $\cos\theta$ will be treated as a value between 0 and 1 with a non-flat distribution that depends on $E_0$ as we will specify later in this Section.

The adopted distribution of $\gamma$ and $\epsilon$ used in this work can be separated in two categories. Each sub-sample is fitted with both categories and we keep the results from the best fit. The first category (r type) works well with distributions of galaxies with low projected $b/a$ values. It comprises the sum of 10 gaussians with a fixed dispersion of 0.08, and mean values going from  0.04 to 0.4 with a step of 0.04 for $\gamma$. For $\epsilon$ we use the same distribution used in PS08, a single log-normal with variable mean and dispersion.

The second category (n type) works well for samples of galaxies with larger $b/a$. The $\gamma$ distribution is the sum of 10 gaussians, with fixed dispersion of 0.08, and means going from 0.1 to 0.82. The $\epsilon$ distribution is that of SSG10, and consists of the sum of the positive side of 10 gaussians centred in 0, with dispersions ranging from 0.02 to 0.2 with a step of 0.02.

\subsection{The dust extinction model}
Following PS08, the model for dust obscuration in this work assumes that the extinction is proportional to the path length of the light through the galaxy. Then, the extinction by dust increases with the inclination of a galaxy. \cite{shao}, \cite{ryden3} and \cite{maller} find that the optical depth increases monotonically with inclination angle.

The model used by PS08 considers an oblate galaxy, with axis ratios $x=B/A$ and $y=C/B$. The total dust extinction as a function of inclination $\theta$ is given by
\begin{equation}
  E(\theta)= \begin{cases}
    E_0(1+y-\cos\theta) & \textmd{if } \cos\theta>y\\
    E_0  & \textmd{if } \cos\theta<y,
  \end{cases}\label{eq:extinction}
\end{equation}
where $E_0$ is the dust extinction in magnitudes in a galaxy edge-on, and $y$ can be extracted from the distribution of $\gamma$. For the reddening, we can assume an analogue model.
\begin{equation}
  R(\theta)= \begin{cases}
    R_0(1+y-\cos\theta) & \textmd{if } \cos\theta>y,\\
    R_0  & \textmd{if } \cos\theta<y,
  \end{cases}\label{eq:reddening}
\end{equation}
where $R_0$ is the edge-on reddening in magnitudes. In the optically thin case, $R_0$ is related to $E_0$ via $R_0=E_0/2.77$ for the $r$ band and the $g-r$ colour. PS08 found that their results do not strongly depend on the proportionality between $E_0$ and $R_0$, and can be applied to both the optically thin and thick cases.

The extincted luminosity function is given by
\begin{equation}
  \phi_{E}(M,\theta)=\phi(M+E(\theta)), \label{eq:lf}
\end{equation}
where $\phi(M)$ is the unextincted luminosity function, this can be calculated using only face-on galaxies from the sample.

PS08 defines the ratio between number density of the observed galaxies (considering extinction) and the unextincted galaxies of a given luminosity as $f_E(M)=\phi_E(M)/\phi(M)$. For reddening PS08 they define $f_R(g-r)$, as the ratio between the reddened and the intrinsic colour distributions.

The ratio between the number of galaxies observed at a given inclination and the intrinsic number of galaxies at a given inclination is calculated multiplying the effects of reddening and extinction together.
\begin{equation}
  \psi(\theta)=\frac{\int_{-\infty}^{\infty}\int_{-\infty}^{\infty} f_E (M)f_R (C) \phi_s(M) \phi_s(C) W(C,M) \textmd{d}C\textmd{d}M}{\int_{-\infty}^{\infty}\int_{-\infty}^{\infty} \phi_s(M) \phi_s(C) W(C,M) \textmd{d}C\textmd{d}M} \label{eq:thetadist}
\end{equation}
Here, $C=g-r$, and $W$ is the correlation between $C$ and $M_r$. PS08 assume that $W$ is Gaussian, and their mean and dispersion can be extracted directly from the data. The sub-index $s$ indicates that the luminosity and colour function correspond to a particular sub-sample of galaxies. $\psi(\theta)$ depends not only on the extinction and reddening, but also on the colour and magnitude range of every sub-sample.

Using the $\psi(\theta)$ distribution, we calculate an inclination $\theta$ for the galaxies in the simulated catalogue, which is used in equations \ref{eq:V} and \ref{eq:W}, and then these results are used in equation \ref{eq:ba} to calculate the projected $b/a$ of each simulated galaxy.

In general, the dust decreases the number of galaxies edge-on relative to face-on, that is, it tends to decrease the number of galaxies with small $b/a$.

\subsection{Extinction and luminosity functions}

\begin{figure}
  \begin{center}
    \includegraphics[width=.44\textwidth]{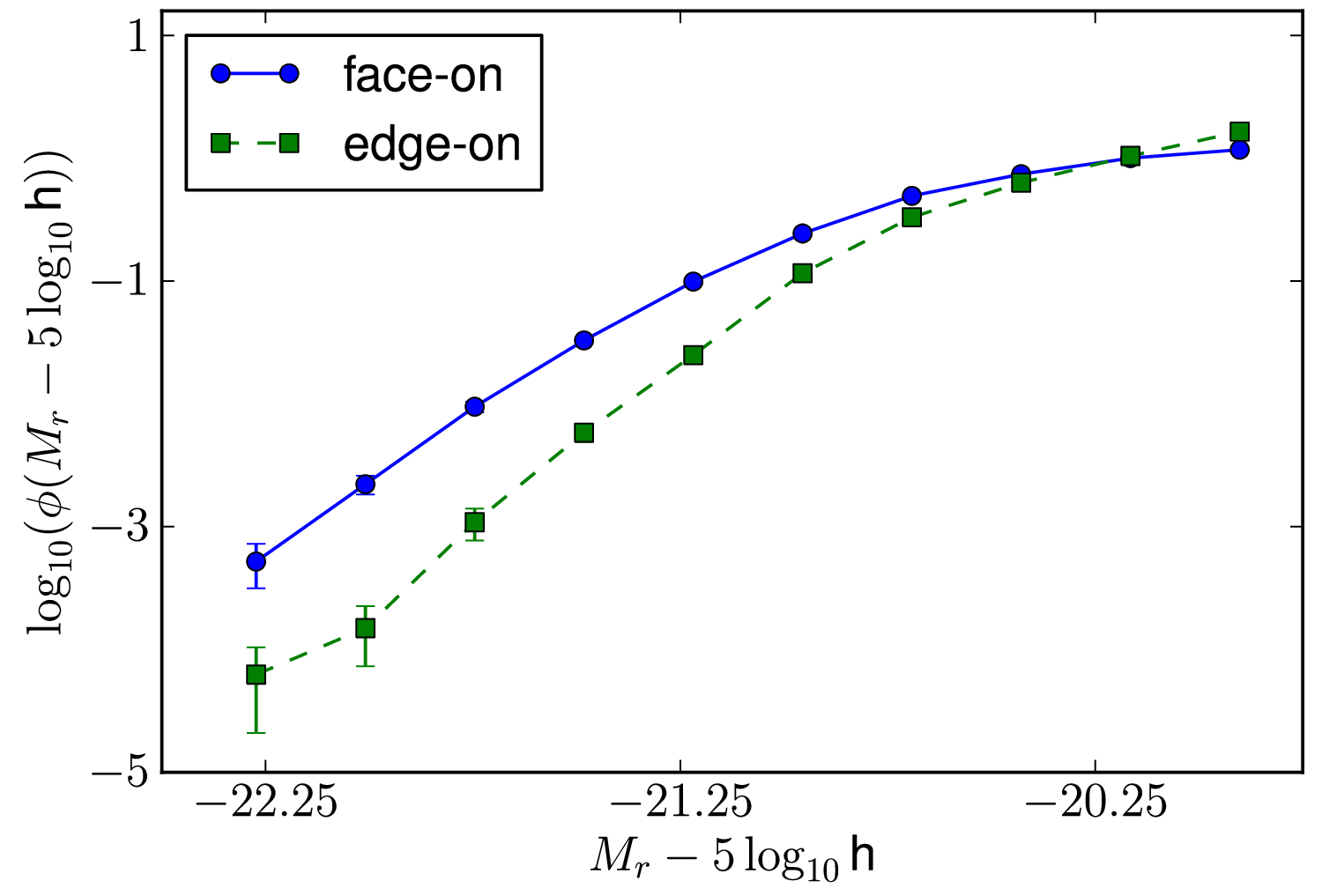}
    \caption{Luminosity function of the face-on and edge-on spirals.\label{fig:LF_exp}}
  \end{center}
\end{figure}

To obtain the value of $E_0$, we use the difference between the luminosity function of the face-on and edge-on galaxies; as an example
Figure \ref{fig:LF_exp} shows the luminosity function of the face-on and edge-on galaxies in the total sample of spirals. There is a clear difference between the two LFs which we will  use to estimate $E_0$ 
following the steps we describe below.

\begin{itemize}
  \item For a given sub-sample of galaxies, we select the 10\% of the galaxies with the highest $b/a$ values (face-on), and the 10\% of galaxies with the lowest $b/a$ (edge-on).   We calculate the LF for each of these selections of galaxies in different luminosity bins.
  \item For each luminosity bin in the face-on LF, we calculate the  difference in magnitudes, to the edge-on LF at the same space density. The average of this value is called $\Delta E$.
  \item We calculate the value of $\langle b/a \rangle$ for the face-on and edge-on galaxies. With this value we get an equivalent $\cos \theta$ value from Table 8 of PS08. We also use the values of $\langle \gamma \rangle$ and $\langle \epsilon \rangle$ from PS08 to obtain $y$.
  \item Using equation \ref{eq:extinction} we calculate the value of $E_0$ corresponding to this $\Delta E$ and the values of $\cos \theta$ and $y$.
  \item We use this $E_0$ to find a new distribution of intrinsic shapes $\gamma$ and $\epsilon$ for the sub-sample.
  \item Then we repeat the calculation of $E_0$ from $\Delta E$, but this time using the values of $\gamma$, $\epsilon$ and the relationship between $\cos \theta$ obtained from the new fit.
  \item Finally we use this value of $E_0$ to obtain the final distributions of $\gamma$ and $\epsilon$.
\end{itemize}

We calculate the differences between the first value of $E_0$, using data from the PS08 fit, and the $E_0$ found using data from our initial fit. We obtain differences that range from 17\% to  101\% between the two values.  We then repeat the measurement and find new values for the extinction ($E_1$) from the fit obtained using the $E_0$ value from the first iteration. The differences between $E_1$ and $E_0$ range from  0.9\% to 16\%. Given these results we decide that one iteration is enough to obtain a converged $E_0$ value.

In the calculation of the error in $E_0$, we only take into account the LF errors for the face- and edge-on galaxies. We do not take into count the errors in the $b/a-\cos\theta$ relation.  We find that this second source of uncertainty  does not affect the most probable value of $E_0$ and only slightly changes its estimated error.  We estimate that this error is always $\leq$ 30\% of the value of $E_0$.

In the case of the sub-samples separated by luminosity we use the $E_0$ value obtained at the median of the magnitude corresponding to each sub-sample. 

\subsection{Metropolis-Hastings algorithm}

To find the distributions of $\epsilon$ and $\gamma$ which present the best fit to the observations we use the Metropolis-Hastings algorithm \citep{met,hast}, which is a Monte Carlo Markov Chain method for optimization.

The Metropolis-Hastings algorithm uses a proposal density ($Q(\vec{x_1},\vec{x})$, where $\vec{x_1}$ and $\vec{x}$ are vectors in the parameter space), to determine a candidate to the next point $\vec{x_1}$ from the current point $\vec{x}$. The probability $\alpha$ that $\vec{x_1}$ is the next point in a chain is given by
\begin{equation}
  \alpha=\frac{F(\vec{x_1})Q(\vec{x_1},\vec{x})}{F(\vec{x})Q(\vec{x},\vec{x_1})}. \label{eq:met-hast} 
\end{equation}
Here, $F(\vec{x})$ is the function we need to maximize. If $Q$ is a symmetric function, then $Q(\vec{x_1},\vec{x})=Q(\vec{x},\vec{x_1})$ and the algorithm is a Metropolis algorithm, with a probability $\alpha$ given by $\alpha=F(\vec{x_1})/F(\vec{x})$.

In this particular case, $\vec{x}$ is the list of percentage of galaxies for each fixed sub-distribution of $\gamma$ and $\epsilon$, for a total of 20 parameters (fits of type n) or 12 parameters (type r). For the proposal density $Q$, we use a Gaussian function, centred in the present point in the chain. Following PS08 we maximize $1/\chi^2$, given by
\begin{equation}
  \chi^2(\vec{x})=\sum_{b/a\,\,bins} \left[\frac{N_{model}(b/a,\vec{x})-N(b/a)}{\sigma_{jackknife}(b/a)}\right]^2, \label{eq:chisquare}
\end{equation}
where $N(b/a)$ is the observed normalized frequency of a given $b/a \pm \Delta b/a$ (with $\Delta b/a$ a half of the bin size), $N_{model}(b/a,\{p\}_i)$ is the normalized distribution given by the model, and  $\sigma_{jackknife}(b/a)$ is the error in the observed distributions for a given $b/a$ calculated using the jackknife method using $10$ jacknife samples.
Throughout this work we use a bin size $2\Delta (b/a) = 0.025$.

\section{The intrinsic shape of SDSS galaxies}
\label{sec:result1}

\begin{figure*}
  \begin{center}
    \includegraphics[width=.85\textwidth]{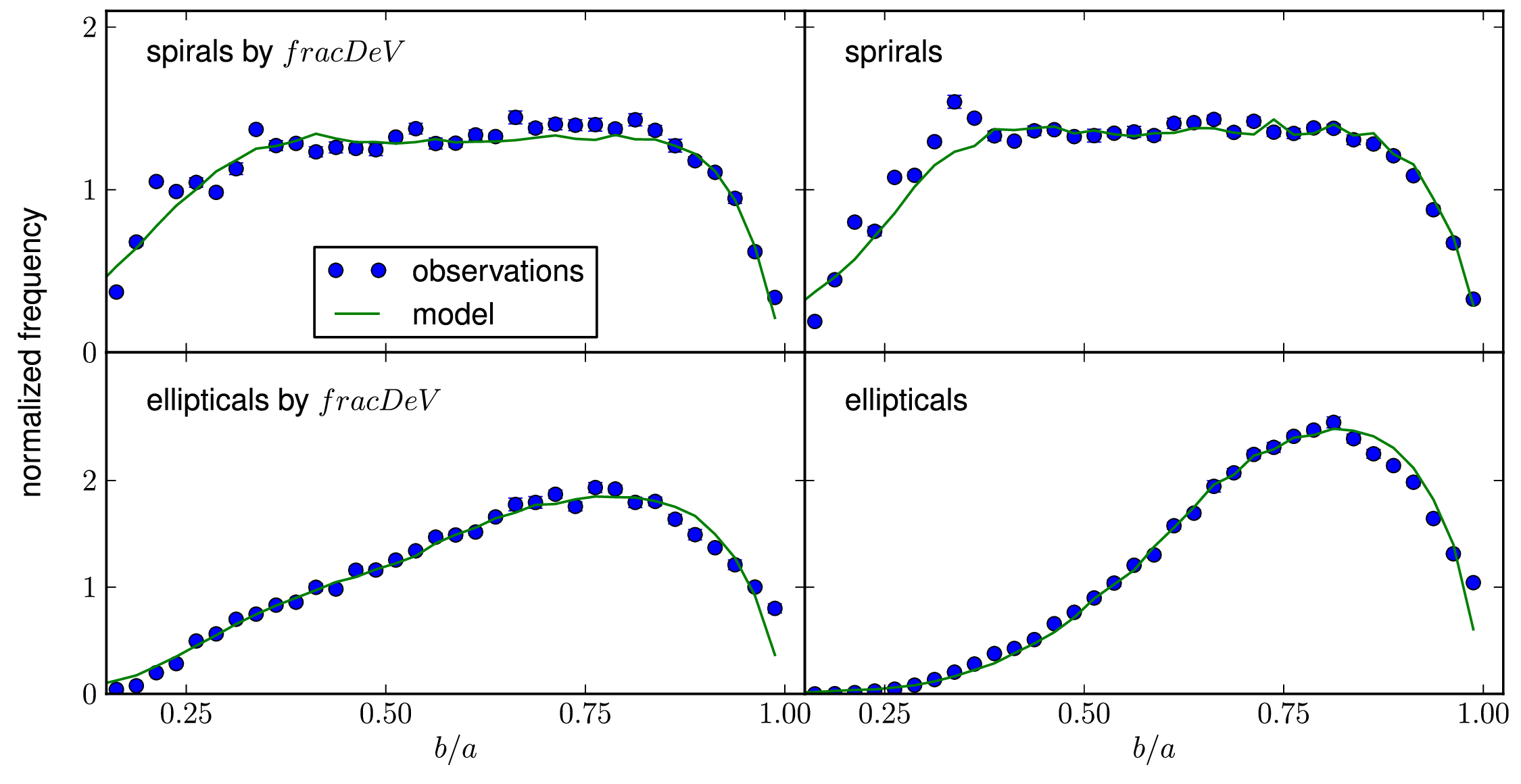}
    \caption{Best fit model $b/a$ distributions compared to the observations. Top: Spirals. Bottom: Ellipticals. Left: Selected only by $fracDeV$. Right: Selected by Galaxy Zoo morphology and $fracDeV$.\label{fig:best}}
  \end{center}
\end{figure*}

Figure \ref{fig:best} shows the best fit models (lines) for the samples selected with and without using Galaxy Zoo morphologies, for spirals and ellipticals (observations are shown as symbols). In the case of the spirals selected only by $fracDeV$ the $E_0$ value used for the fit was the value from PS08. The results from these fits and their comparison to PS08 will be discussed in the following sections. The models used in the fits are type r for the two spiral samples, and type n for the two elliptical samples (throughout, all the elliptical sub-samples are fitted using type n).

\subsection{Spirals}
\begin{figure}
  \begin{center}
    \includegraphics[width=.44\textwidth]{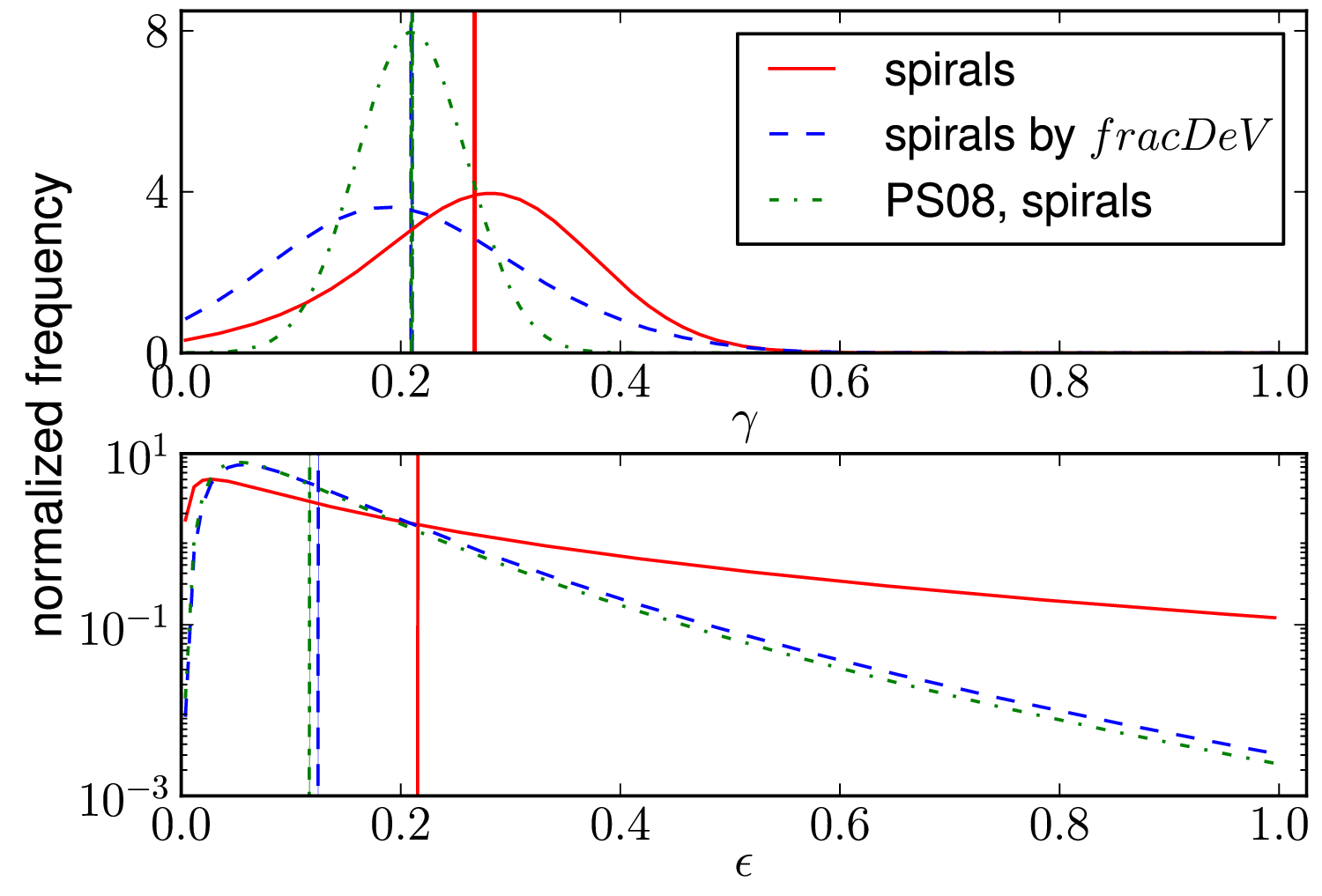}
    \caption{Top: Distribution for $\gamma$ in spirals selected by $fracDeV$ and Galaxy Zoo morphology. Bottom: Distribution for $\epsilon$ in spirals.  We also show the results for PS08 and the results for spirals selected only by $fracDeV$. Vertical lines show the mean of $\gamma$ and $\epsilon$ of each sample.\label{fig:gam-eps}}
  \end{center}
\end{figure}

Figure \ref{fig:gam-eps} shows the $\gamma$ distribution and the $\epsilon$ distribution obtained for these samples, as well as the best fit from PS08, who used a single gaussian function to model the $\gamma$ distribution and a log-normal distribution to model the $\epsilon$ distribution.
For the sample separated by Galaxy Zoo and $fracDeV$ the dust extinction is $E_0=0.284^{+ 0.015}_{-0.026}$, whereas the extinction found by PS08 for spirals is slightly higher although consistent with our estimate, $E_0=0.44 \pm 0.24$.
The figure shows that the result obtained by the new non-parametric model is in good agreement with the result found in PS08 for spirals selected by $fracDeV$. The mean of the $\gamma$ distribution from the sample of spirals selected by $fracDeV$ is $0.21 \pm 0.008 $, with a dispersion $ 0.108 \pm 0.004 $, very similar to the PS08 results ($\langle\gamma\rangle=0.21\pm0.02$ and $\sigma_{\gamma}=0.05\pm0.015$). But the shape we obtain for the $\gamma$ distribution is not well represented by Gaussian distribution (the mean does not match the mode), allowing more galaxies with larger minor to major axis ratio, as well as flatter discs. The $\epsilon$ distribution does not show significant differences between the two samples.

Regarding the samples of spirals selected with and without adding Zoo morphologies, the  $\gamma$ distribution tends to have a larger minor to major axis ratio when Zoo morphologies are used, with a mean given by $ 0.267 \pm 0.009 $ and with a small dispersion, given by  $0.102 \pm 0.004$. The $\epsilon$ distribution shows that galactic disks are rounder for galaxies selected only by $fracDeV$, with a $\langle \epsilon \rangle$ given by $ 0.125 \pm 0.008 $, and a dispersion given by $0.1 \pm 0.012$. In comparison, $\langle \epsilon \rangle=0.22 \pm 0.013$ for the sample selected by $fracDeV$ and Galaxy Zoo (In PS08, we have $\langle\ln\epsilon\rangle=-2.33 \pm 0.13$, and $\sigma_{\ln \epsilon}=1.4 \pm 0.1$, which correspond to $\langle \epsilon \rangle = 0.18 $ and $\sigma_{\epsilon} = 0.12 $).

\subsection{Ellipticals}

PS08 report the possible contamination by spiral galaxies in a sample of elliptical galaxies selected by $fracDeV$. The distribution of $b/a$ from ellipticals separated by $fracDeV$ (bottom left panel on Figure \ref{fig:best}) shows a ``hump'' around $b/a= 0.2$ to $0.4$, which disappears in the sample selected using Galaxy Zoo morphology and $fracDeV$ (bottom right panel on Figure \ref{fig:best}). This can be caused by the presence of contamination of spirals, since the hump at such values of $b/a$ is more easily produced by flat discs.

\begin{figure}
  \begin{center}
    \includegraphics[width=.44\textwidth]{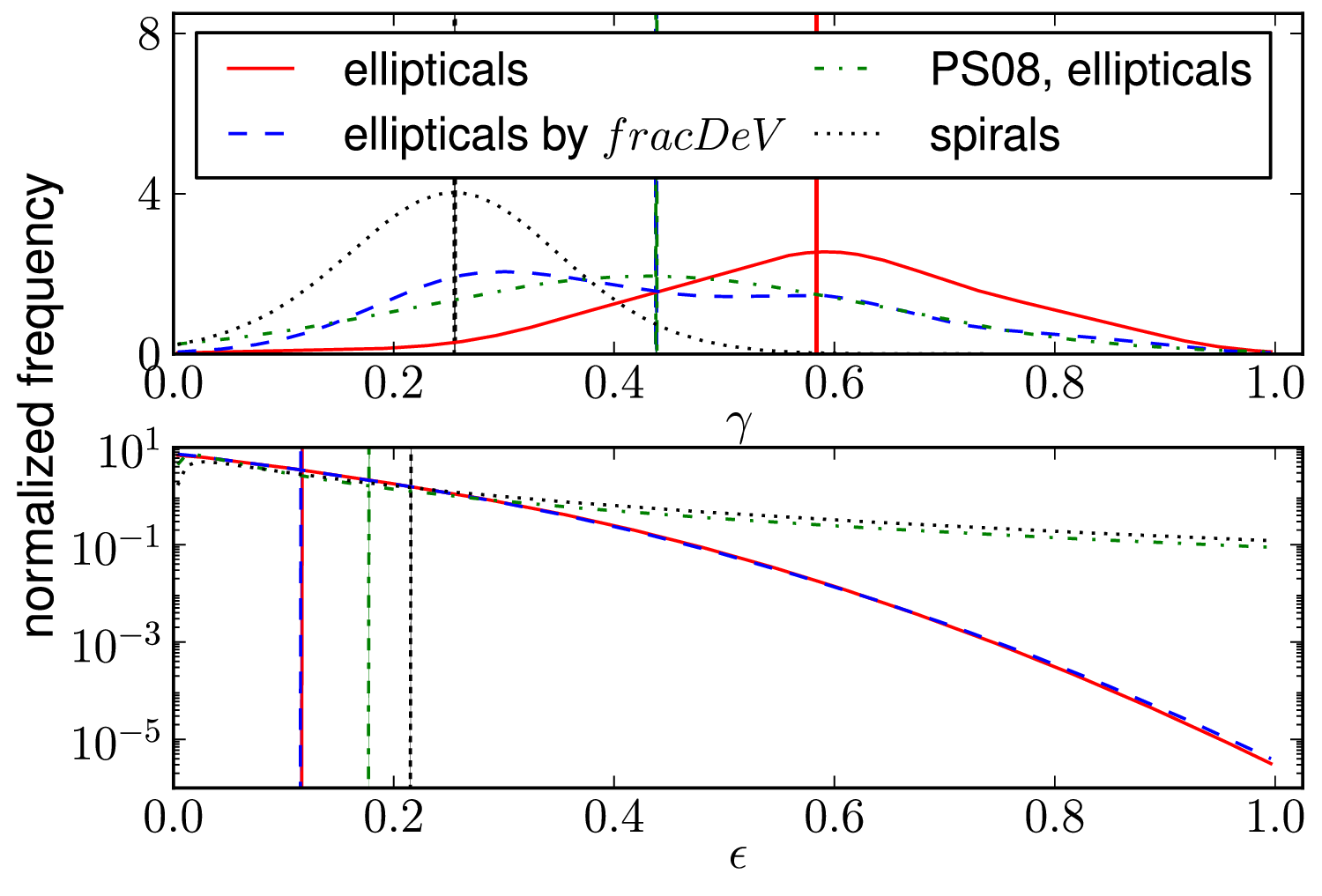}
    \caption{Top: Distribution for $\gamma$ in ellipticals selected by $fracDeV$ and Galaxy Zoo morphology. Bottom: Distribution for $\epsilon$ in ellipticals.  We also show the results for PS08, the results for ellipticals selected only by $fracDeV$, and the distribution for spirals selected by $fracDeV$ and Galaxy Zoo morphology. Vertical lines shows the mean of $\gamma$ and $\epsilon$ for each sample.\label{fig:gam-eps2}}
  \end{center}
\end{figure}

Figure \ref{fig:gam-eps2} shows the $\gamma$ and $\epsilon$ distribution of the result obtained by PS08. The $\gamma$ distribution shows that the results using this method are in good agreement with the results from PS08. The mean of the $\gamma$ distribution for the sample of ellipticals selected using only $fracDeV$ obtained with the method used in this work is $0.438 \pm 0.005 $ , and the dispersion is $ 0.196 \pm 0.005 $, similar to the PS08 results ($\langle\gamma\rangle=0.43\pm0.06$, $\sigma_{\gamma}=0.21\pm0.02$). But the $\gamma$ distribution obtained by this work shows a peak at $\gamma$ below 0.5, close to the peak shown in the $\gamma$ distribution for spirals (also shown in the figure to reinforce this result); this peak allows the new model to fit the hump discussed before. The model used by PS08 is not able to show the presence of discs in the sample but averages them with the intrinsic shapes of the ellipsoids, slightly biasing the results. The $\epsilon$ distribution shows that the shapes of the galaxies in this model are rounder than for the galaxies using the PS08 model.

The $\gamma$ distributions for ellipticals selected using Galaxy Zoo morphology and $fracDeV$ show that the elliptical galaxies in this sample have a minor to major axis ratio larger than the sample selected using only $fracDeV$, with a mean given by $ 0.584 \pm 0.006 $. Also, the peak on the $\gamma$ distribution for the sample selected by Galaxy Zoo data is unlikely to be caused by discs, but the peak on the $\gamma$ distribution for the sample selected only by $fracDeV$ can be caused by spiral contamination in the sample. This supports the idea that the ``hump'' in the bottom left panel in Figure \ref{fig:best} is caused by contamination from spirals. The $\epsilon$ distribution does not show significant differences between the two samples.

The $\gamma$ distribution shows that spirals are flatter than ellipticals. The $\epsilon$ distribution shows that the spiral discs are more elliptical than would be face-on ellipticals.

\subsection{Low and High $fracDeV$ discard galaxies}

The results presented in this section pose questions about the nature of galaxies with exponential luminosity profiles that failed to be classified as spirals by Galaxy Zoo volunteers.  Figure \ref{fig:gam-eps} shows that the final full sample of spirals has larger values of $\gamma$ and $\epsilon$ compared to the results of PS08, as well as to our fit of the comparison sample. This can lead to the conclusion that the low $fracDeV$ galaxies discarded by the Galaxy Zoo classification are thinner and have a rounder discs than our final sample of spirals.

\begin{figure}
  \begin{center}
    \includegraphics[width=.44\textwidth]{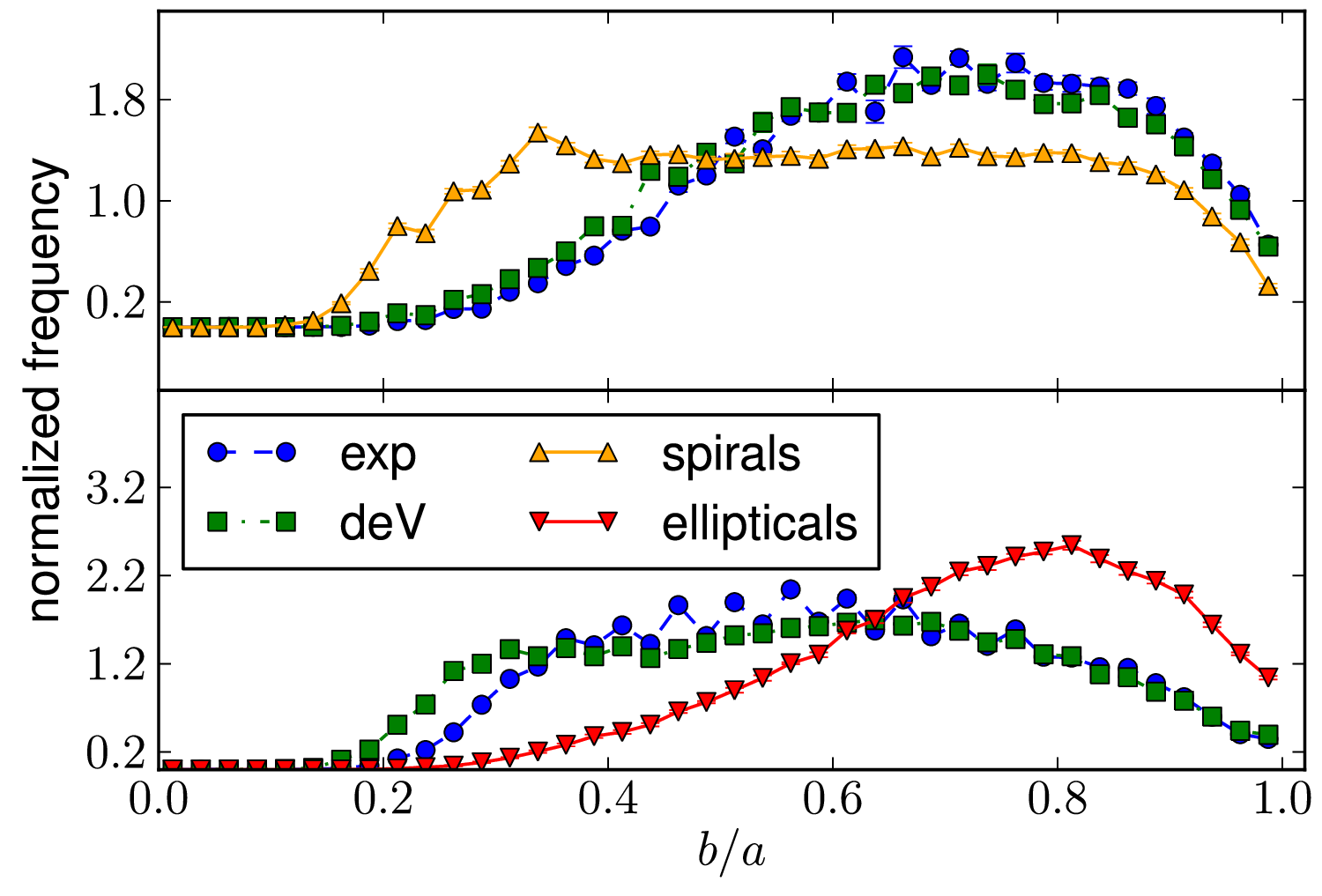}
    \caption{$b/a$ distribution for the low and high $fracDeV$ discard galaxies due to their Galaxy Zoo classification. The exp label indicates a $b/a$ calculated assuming an exponential profile. The deV label indicates that $b/a$ was obtained assuming a de Vaucouleurs profile. Top: $b/a$ distribution of the low $fracDeV$ discard galaxies, along with the $b/a$ distribution of our sample of spirals. Bottom: $b/a$ distribution of the high $fracDeV$ discard galaxies, along with the distribution of our sample of ellipticals.\label{fig:ba_noexp-deV}}
  \end{center}
\end{figure}

The median values of the discard low $fracDeV$ galaxies of magnitude, colour and size are $-20.43$, $0.77$ and $2.84$ kpc, respectively. According to Table \ref{table:not_exp}, these galaxies are similar to the other spiral galaxies in brightness and colour, but smaller than the average spiral. The votes in Galaxy Zoo indicate that very few of these galaxies are classified as mergers or star/artefacts (4.39 \% and 0.75 \%, respectively) whereas a considerable fraction of them are classified as ellipticals (36.11 \%).  However, their majority is not given a definite classification, that is, the citizen scientists of Galaxy Zoo cannot reach a consensus about their morphology. 
Figure \ref{fig:ba_noexp-deV} shows the $b/a$ distribution of the low $fracDeV$ discard galaxies. Given that we do not know the true morphology of these galaxies, we show the distributions resulting from using the two estimates of model $b/a$, the one obtained assuming an exponential profile and the other assuming a de Vaucouleurs profile. These distributions indicate that, contrary to the expected results, these galaxies are rounder than the average spiral, with a $b/a$ distribution more similar to that of elliptical galaxies.  How is it possible that after removing these galaxies (with apparently high $\gamma$) we obtain higher values of $\gamma$ for their intrinsic shapes?  The answer lies in the dust content.

PS08 measured the dust directly from the fit to the distribution of apparent galaxy shapes.  They took advantage of the fact that dust affects the $b/a$ distribution by decreasing the number of galaxies with low $b/a$.  When PS08 estimate the $\gamma$ and $\epsilon$ their algorithm assumes that the dust content is the cause of the number galaxies with low $b/a$, and infers that galaxies are thinner. However, if the dust content were lower the galaxies could actually be intrinsically thicker. In our case, we used the dust estimated by PS08 to fit the $b/a$ distribution of the comparison sample, and therefore we obtained consistent conclusions with PS08 for this sample.   
However, when we use the edge- and face-on luminosity function method to estimate dust
for the low $fracDeV$ discard galaxies we obtain $E_0=0.145^{+0.053}_{-0.046}$, which is actually lower than that of the main spiral sample.  Therefore, the discards have very low dust and are indeed rounder than the galaxies in our spirals sample.

The galaxies with a high $fracDeV$ but discarded by Galaxy Zoo are different. Since dust does not play a role in the determination of the shape of ellipticals, the difference in the results found by PS08 and ours are consistent with the $b/a$ distribution of the high $fracDeV$ discard galaxies (Figure \ref{fig:ba_noexp-deV}). This distribution is more similar to that of spirals than of ellipticals, with an important number of galaxies with low values of $b/a$. Their medians in magnitude, colour and size are $-20.513$, $0.91$ and $2.88$ kpc, respectively. These are consistent with slightly blue ellipticals or very red and small spirals (Tables  \ref{table:not_exp} and \ref{table:not_deV}). The fractions of Galaxy Zoo votes show that the majority of galaxies in this sample (61.06 \%) are classified as spirals, and the rest are galaxies without a definite classification from Galaxy Zoo, with a small part of mergers and star/artefacts (less than 3.3 \% combining both types). Using our method to estimate the dust of this sample (using the values of spirals), we find a value of $E_0=0.22^{+ 0.039}_{-0.027}$, very similar to that of the full sample of spirals.

\section{The intrinsic shape of SDSS galaxies with Galaxy Zoo morphologies}
\label{sec:result2}

\begin{figure*}
  \begin{center}
    \includegraphics[width=.85\textwidth]{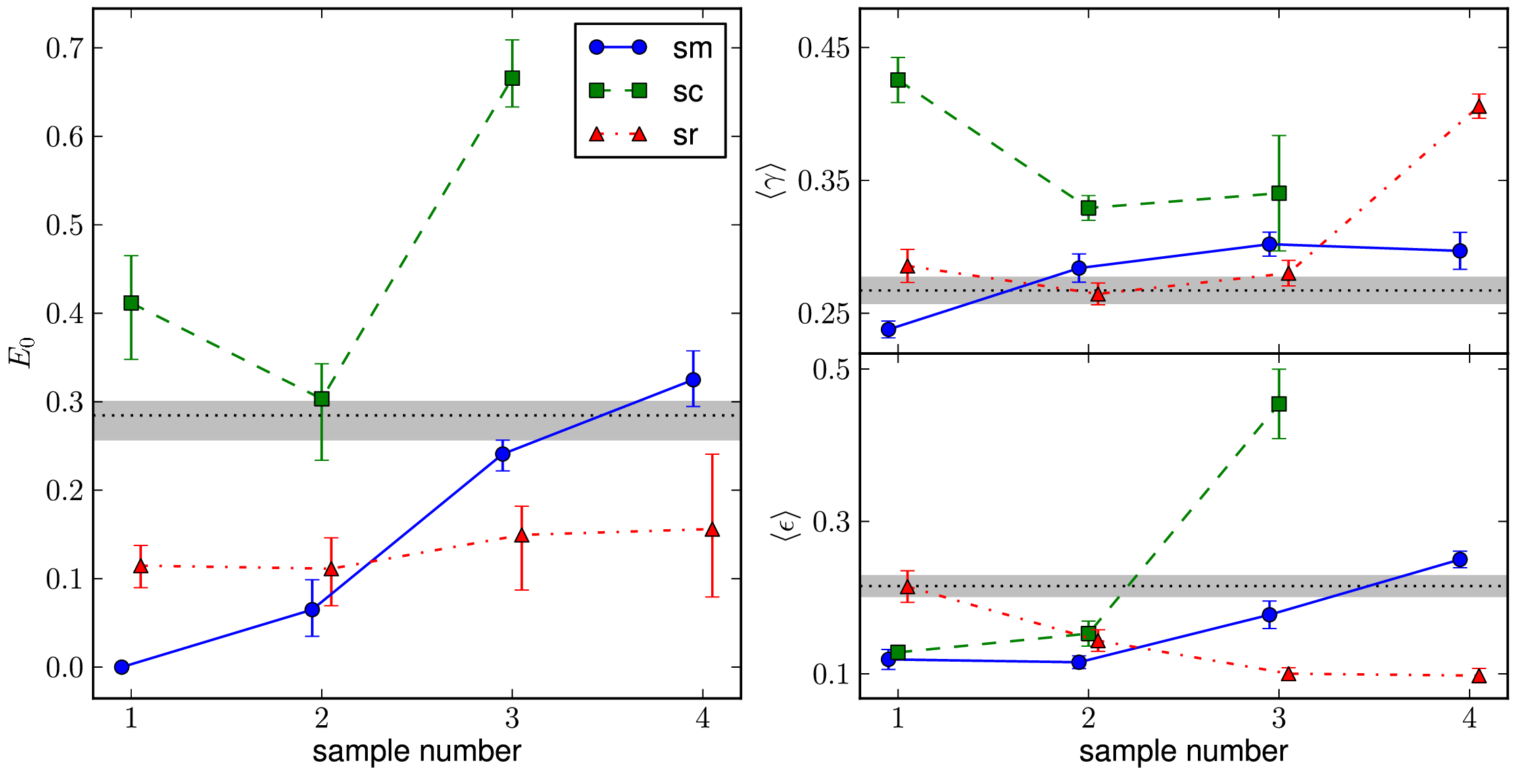}
    \caption{Best fit result for spirals, separated by different properties. The x-axes run with the sample number. The solid lines with circles indicate the samples separated by luminosity, where the luminosity increase with the sample number. The dashed lines with squares represent the samples separated by colour, where the value of $g-r$ increases with the sample number. The dash-dotted lines with triangles are the samples separated by size, where the size increases with the sample number. The dotted line and the shadowed area indicate the value and associated error (respectively) of the total sample of spirals. In the sm and sr samples a small shift in the horizontal direction has been introduced to improve clarity. Left: Variation of $E_0$ with galaxy properties. Top right Variation of $\langle \gamma \rangle$. Bottom right: Variation of $\langle \epsilon \rangle$. \label{fig:exp_int}}
  \end{center}
\end{figure*}

In this Section, we discuss the results for the intrinsic shapes of galaxies in different sub-samples, selected by their intrinsic properties.  Figure \ref{fig:exp_int} shows the results for the mean values of $\gamma$, $\epsilon$ and $E_0$ for the different sub-samples of spirals selected according to their luminosity, size and colour, as presented in Table \ref{table:not_exp}. Table \ref{table:texp_tot} shows the values that characterize the $\gamma$ and $\epsilon$ distribution, along with the $\chi^2/\textmd{d.o.f.}$ and, for spirals, the value of $E_0$ and the type of fit (for ellipticals all the fits are type n).

\begin{figure}
  \begin{center}
    \includegraphics[width=.44\textwidth]{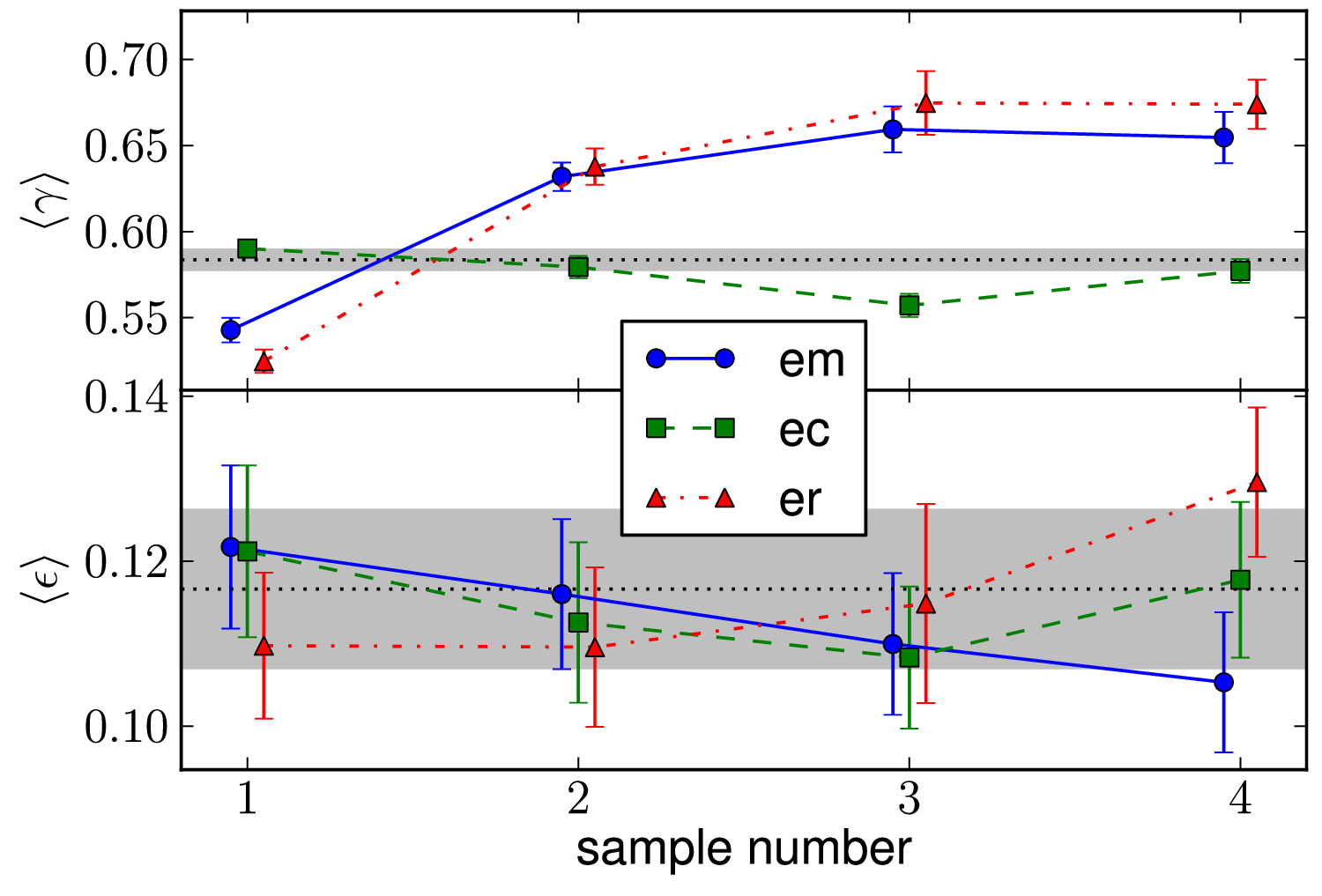}
    \caption{Best fit result for ellipticals, separated by different galaxy properties. Symbols and line types represent the same as in Figure \ref{fig:exp_int} for elliptical samples. Top: Variation of $\langle \gamma \rangle$. Bottom: Variation of $\langle \epsilon \rangle$. \label{fig:deV_int}}
  \end{center}
\end{figure}

Figure \ref{fig:deV_int} shows the results for the mean values of $\gamma$ and $\epsilon$ for ellipticals defined in Table \ref{table:not_deV} separated by their intrinsic characteristics. It is important to bear in mind that the dust extinction for the elliptical samples is set to 0.

These results are discussed in the following subsections, for spiral and elliptical galaxies separately.

\subsection{Dependence on magnitude}

The samples labelled as sm in Figure \ref{fig:exp_int} shows that the brighter spiral galaxies have a larger dust extinction than fainter ones, with a clear increasing trend.

The $\langle \gamma \rangle$ values show that the faintest spiral galaxies tend to have smaller minor to major axis ratio than the brightest ones, whereas $\langle \epsilon \rangle$ values show that the brightest galaxies have less circular discs than the fainter ones. There is a clear relationship between ellipticity and magnitude for spirals.

The $\langle \epsilon \rangle$ values of the samples labelled as em in Figure \ref{fig:deV_int} show that the more luminous elliptical galaxies tend to be rounder in $B/A$ than the less luminous ones, although the variations are within the error bars.

The $\gamma$ distributions show that in general the more luminous elliptical galaxies have larger minor to major axis than the less luminous ones. The more luminous ellipticals tend to be more spherical than the fainter elliptical galaxies, which have more irregular shapes. The mean in the values of $\gamma$ in Table \ref{table:texp_tot} and Figure \ref{fig:deV_int} show an important decrease in the mean from brighter to fainter galaxies.

\subsection{Dependence on colour}

The sub-sample named sc4, i.e. the reddest spiral galaxy sub-sample, is not shown in Figure \ref{fig:exp_int} (the $b/a$ distribution of this sub-sample is in Figure \ref{fig:ba-fordist}) because non of our models are able to fit the observed $b/a$ distribution of this sub-sample.  Sub-sample sc3 is also difficult to fit, except when using the method by PS08 (for sc4 this was also not possible).  This is shown in Table \ref{table:texp_tot} which includes a column with the fitting method used. Also, it is important to take into account that in the calculation of the errors in the shape parameters for this sample (calculated with the jackknife method, as all the errorbars in this work) it was necessary to remove 6 outlier jackknifes (from a total of 100) by performing an interquartile range technique (this technique did not remove any jackknife results from the analysis of any other subsample). There were no outliers in the results of $E_0$ from the jackknife samples.

Figure \ref{fig:exp_int} shows that the sc sub-samples exhibit a clear variation of the extinction with colour. $E_0$ varies from 0.4 magnitudes for the bluest sample, to $\sim$ 0.7 for the reddest sample. The value of $\langle \gamma \rangle$ also varies with colour; the bluer sample tends to show larger $\gamma$ values than redder ones. The value of $\langle \epsilon \rangle$ is highest for the reddest sample. This trend is still present even when including the outliers for sc3, and shows that the bluer spirals have more spherical shapes. 

The $\langle \epsilon \rangle$ values of the elliptical ec samples in Figure \ref{fig:deV_int} do not show a great variation with colour.
This is also the case for the values of $\langle \gamma \rangle$. This is confirmed by the top right panel of Figure \ref{fig:ba-fordist2}, where the $b/a$ distributions of all the sub-samples of ellipticals separated by colour are very similar.

\subsection{Dependence on size}

The $b/a$ distribution of spiral galaxy samples corresponding to different sizes (labelled sr in Figure \ref{fig:exp_int}) are fit by models with slightly higher amounts of dust extinction for bigger galaxies. But this trend is within the error bars. 

The value of $\langle \gamma \rangle$ is higher for larger galaxies, indicating a thicker disc for larger spirals. The value of $\langle \epsilon \rangle$ varies significantly with the galaxy size, indicating that the larger galaxies tend to have a rounder shapes than the smaller galaxies.
 
The $\langle \epsilon \rangle$ values for the samples labelled er in Figure \ref{fig:deV_int} indicate that the sample of small ellipticals are characterized by higher ellipticities than larger ones.  For the $\langle \gamma \rangle$ values, the larger ellipticals tend to be rounder than small galaxies. 

\subsection{Interdependence between parameters}
\label{ssec:interdependence}

Galaxy luminosity, colour and size are correlated such that brighter galaxies tend to be redder and larger.  Consequently, the trends found in this Section are not due to changes in one individual property alone.  
For instance, the spiral galaxy samples divided by galaxy luminosity as presented in Table \ref{table:not_exp} are characterized by median colours that differ by up to $0.06$ magnitudes.  In the case of size, the colour difference between the smallest and largest galaxies is $0.05$ magnitudes. The colour selected spirals show median absolute magnitudes with $0.13$ magnitudes difference and $0.4$kpc in their sizes.  For elliptical galaxies, the differences are comparable.  Notice, however, that these variations are smaller than the typical separation between the cuts that make our sub-samples (Tables \ref{table:not_exp} and \ref{table:not_deV}), so the effect should be small.

In an ideal situation we would make more stringent cuts in the sub-samples of galaxies so that only one of the three variables changes.  However, the correlation between luminosity and size is very tight and does not allow for this ideal situation to be reached.  It is possible though to make the following new sub-samples. (i)  When selected by colour, we remove galaxies until the medians of size and absolute magnitude are the same as those of the total sample; (ii) in samples selected by luminosity or size, the colour is forced to show a constant median across the samples.  This is the case for both elliptical and spiral galaxy samples.

With these new samples we repeat the analysis of this Section and find, as expected, that the values of $E_0$, $\langle \gamma \rangle$ and $\langle \epsilon \rangle$ are consistent with those for the samples in Tables \ref{table:not_exp} and \ref{table:not_deV}.  The new sub-samples also show the same trends as a function of galaxy properties as in Figures \ref{fig:exp_int} and \ref{fig:deV_int}.  This is mostly due to the small change in the median values with respect to the previous samples; this can also be inferred from the $b/a$ distributions, which do not show important changes with respect to those in Figures \ref{fig:ba-fordist} and \ref{fig:ba-fordist2}.

\section{Discussion and conclusions}
\label{sec:conclusions}

We build upon work by PS08 to obtain the dust extinction and the distribution of the intrinsic shapes of elliptical and spiral galaxies in the SDSS DR8. The new model is an improved version of the one presented in 
PS08 since (i) we now apply the method by SSG10, of modelling the $\gamma=C/A$ and $\epsilon=1-B/A$ distributions as sums of many Gaussian distributions with fixed dispersions and means, with different weights , (ii)  we add the data from the Galaxy Zoo project \citep{lintott2} as a second parameter to define spiral and elliptical galaxy samples in addition to the $fracDeV$ parameter, and (iii) we measure the dust extinction affecting each galaxy sample by studying the face- and edge-on luminosity functions (only for spiral galaxies).

We show that the use of the Galaxy Zoo morphology with $fracDeV$ to classify the galaxies into spirals and ellipticals improves the accuracy of the selection with respect to using the $fracDeV$ parameter alone.  In particular, the use of Galaxy Zoo morphologies helps to eliminate contamination of spirals in samples of ellipticals.  Spiral galaxy interlopers cause the inferred intrinsic shapes of ellipticals to be biased toward flatter shapes.

The improved model shows a good agreement with the results of PS08 for the same samples of ellipticals and spirals selected using only a limit on $fracDeV$, finding similar values for both the mean and dispersion of $\gamma$ and $\epsilon$. However, our improved modelling shows details which the PS08 model was not be able to show, such as the excess of galaxies with low $\gamma$ (disky shapes) corresponding to the spiral galaxy contaminants in the sample of $fracDeV$ ellipticals.

For the sample of spirals selected by both Galaxy Zoo morphology and $fracDeV$, we found a lower value of extinction than for the $fracDeV$ spirals of PS08, of $E_0=0.284^{+0.015}_{-0.026}$ compared to $E_0=0.44 \pm 0.24$ of PS08. This can be due to a contamination of elliptical galaxies in the spiral sample used by PS08. This contamination can cause an excess of galaxies with high $b/a$ values, which the PS08 model compensates by adding more dust \footnote{PS08 determine the value of $E_0$ and the distributions of $\gamma$ and $\epsilon$ directly from the $b/a$ distribution}; as a general rule, more dust implies less galaxies with low $b/a$.

The $\gamma$ distribution for the total sample of spirals shows that the spiral discs are in general thicker than previous estimates, with an average $\gamma = 0.267 \pm 0.009 $. The $\epsilon$ distribution shows that spiral galaxy discs are slightly less round than estimated by PS08. The difference in $\langle \epsilon \rangle$ between the sample separated by $fracDeV$ and the sample separated by $fracDeV$ and Galaxy Zoo is very similar to the one between the PS08 and Galaxy Zoo selected samples, showing that the discrepancy is probably due to some level of contamination of ellipticals in the spirals sample.

The $\gamma$ and $\epsilon$ distributions for the sample of elliptical galaxies selected by Galaxy Zoo morphology and $fracDeV$ show that these galaxies are more spherical than the estimates of PS08. The $\gamma$ distribution has a mean in $ 0.584 \pm 0.006 $.  
The increase in $\langle \gamma \rangle$ is mainly due to the fact that the sample used in this work has less contamination by spirals in the elliptical sample that the samples used in previous works, thanks to the addition of Galaxy Zoo morphologies.
The $\epsilon$ distribution shows that the $B/A$ ratios are smaller for spirals, that is, discs are less round than the typical elliptical galaxy.

\begin{figure}
  \begin{center}
    \includegraphics[width=.49\textwidth]{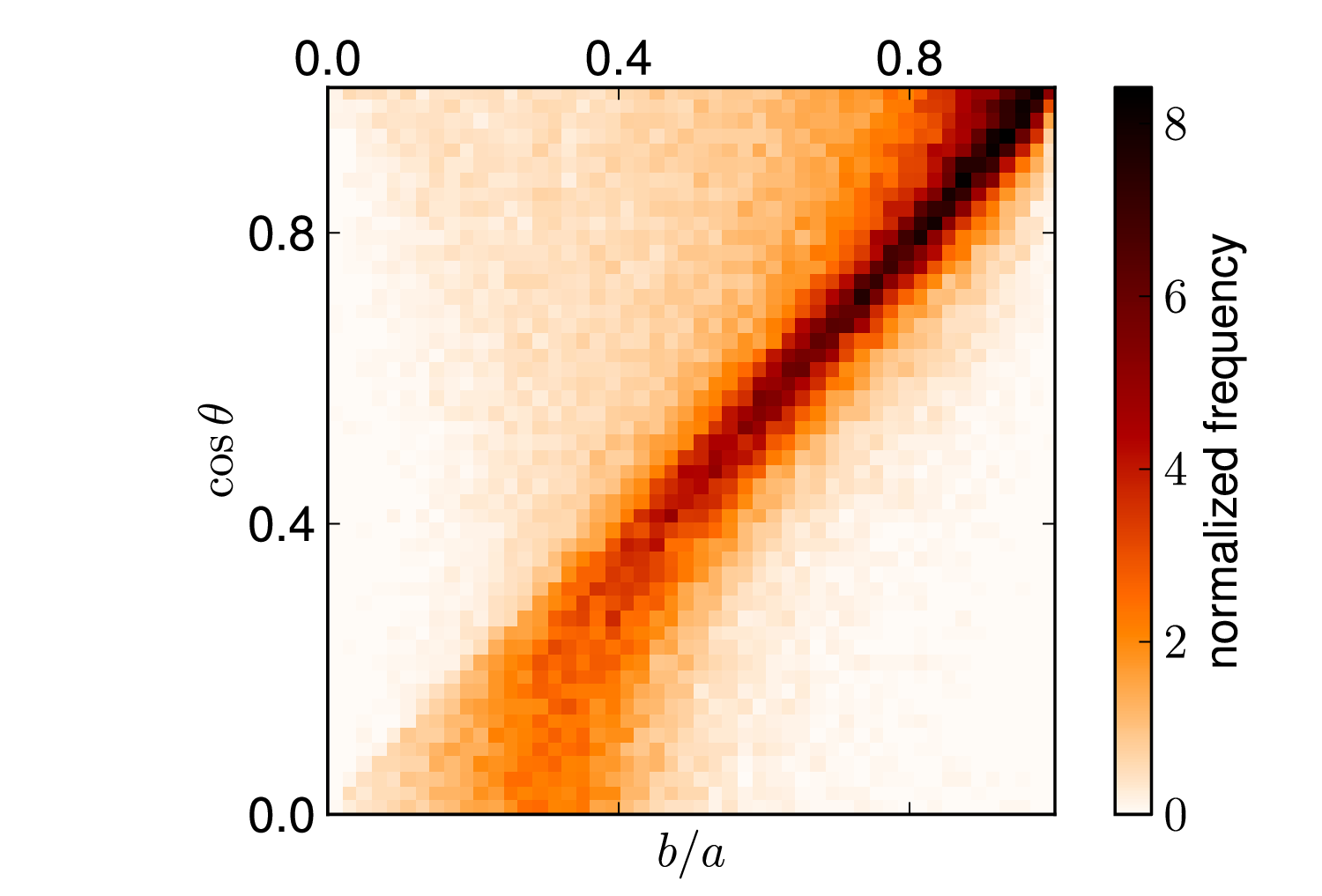}
    \caption{Bi-dimensional distribution of $b/a$ and $\cos \theta$ for the total sample of spirals.\label{fig:angle}}
  \end{center}
\end{figure}

We use these results to calculate the statistical distribution of inclination angle $\theta$ for any given $b/a$. Figure \ref{fig:angle} shows the bi-dimensional distribution of $\cos \theta$ and $b/a$ for the total sample of spirals. This distribution shows that the relation between $\cos \theta$ and $b/a$ is close to a lineal relation, but with an important dispersion.  For small values of $b/a$ or $\cos \theta$, the relation deviates from the linear relation.

We studied the variation of the intrinsic shapes of spirals and ellipticals with absolute magnitude in the r-band, $g-r$ colour, and size. For spirals, we found that the most luminous galaxies tend to have thicker discs than fainter ones. The dust extinction tends to be lower for fainter galaxies.  As a function of $g-r$ colour, bluer galaxies tend to be more spherical than redder ones. The dust extinction shows a clear relation with colour, in which the reddest galaxies have more dust extinction than bluer ones. Galaxies with larger sizes tend to have thicker discs than smaller galaxies. The dust extinction does not show any strong correlation with the size.

For elliptical galaxies, the absolute magnitude shows a strong correlation with the $\gamma$ distribution, so that the more luminous galaxies tend to have larger minor to major axis ratios. The $\epsilon$ distribution shows that the brightest galaxies tend to have larger ellipticities. With $g-r$ colour, the $\epsilon$ and $\gamma$ distributions show no change. With the galaxy size, $\epsilon$ shows a slight trend in which the larger galaxies tend to have a larger $\epsilon$ , but the $\gamma$ distribution shows that small ellipticals are flatter than the larger ones.

These results can be used to test analytical and semi-analytical galaxy formation models, and can give us some clues about the relation between star formation and/or SN feedback and the thickness of galaxy discs, since the intrinsic characteristics that depend on star formation (such as colour and dust content) are clearly related with the shape of a galaxy.

\section*{Acknowledgements}

We want to acknowledge the very useful comments from Michael Strauss, which are an important part of this paper. Also, we want to acknowledge to Claudia Lagos for her comments.

SR want to acknowledge to the people of the SAG group in PUC (Chile), UNLP and UNCOR (Argentina) and UNAM (Mexico), for the comments and the insightful questions, that led to improvements in this work.

SR was supported by Centro de Astronom\'ia y Tecnolog\'ías Afines (CATA) BASAL PFB-06. NP was supported by Proyecto Fondecyt Regular 1110328.

The calculations for this paper were performed in Geryon cluster, at the Centro de Astro-Ingenier\'ia at Pontficia Universidad Cat\'olica de Chile.

Funding for SDSS-III has been provided by the Alfred P. Sloan Foundation, the Participating Institutions, the National Science Foundation, and the U.S. Department of Energy Office of Science. The SDSS-III web site is http://www.sdss3.org/.

SDSS-III is managed by the Astrophysical Research Consortium for the Participating Institutions of the SDSS-III Collaboration including the University of Arizona, the Brazilian Participation Group, Brookhaven National Laboratory, University of Cambridge, Carnegie Mellon University, University of Florida, the French Participation Group, the German Participation Group, Harvard University, the Instituto de Astrofisica de Canarias, the Michigan State/Notre Dame/JINA Participation Group, Johns Hopkins University, Lawrence Berkeley National Laboratory, Max Planck Institute for Astrophysics, Max Planck Institute for Extraterrestrial Physics, New Mexico State University, New York University, Ohio State University, Pennsylvania State University, University of Portsmouth, Princeton University, the Spanish Participation Group, University of Tokyo, University of Utah, Vanderbilt University, University of Virginia, University of Washington, and Yale University.

\begin{table*}
  \begin{center}
    \caption{Means and dispersions in $\gamma$ and $\epsilon$ for sub-samples separated by different characteristic. Top: Spirals. Bottom: Ellipticals.\label{table:texp_tot}}
    \begin{tabular}{cccccccc}
      \hline
      Sample & $E_0$ & $\langle \gamma \rangle$ & $\sigma_{\gamma}$ & $\langle \epsilon \rangle$ & $\sigma_{\epsilon}$ & $\chi^2/\textmd{d.o.f.}$ & Fit type\\\hline
      total & $0.284^{+0.015}_{-0.026}$ & $0.267 \pm 0.009$ & $0.102 \pm 0.004$ & $0.215 \pm 0.013$ & $0.216 \pm 0.008$ & 5.043 & r    \\\hline
      sm1   & $0$                       & $0.238 \pm 0.006$ & $0.107 \pm 0.004$ & $0.119 \pm 0.013$ & $0.099 \pm 0.021$ & 2.318 & r    \\
      sm2   & $0.065^{+0.034}_{-0.03}$  & $0.284 \pm 0.011$ & $0.155 \pm 0.01$  & $0.115 \pm 0.008$ & $0.104 \pm 0.005$ & 4.232 & n    \\
      sm3   & $0.241^{+0.016}_{-0.019}$ & $0.302 \pm 0.009$ & $0.107 \pm 0.004$ & $0.178 \pm 0.018$ & $0.16 \pm 0.02$   & 3.175 & r    \\
      sm4   & $0.325^{+0.032}_{-0.03}$  & $0.297 \pm 0.013$ & $0.105 \pm 0.006$ & $0.25 \pm 0.011 $ & $0.197 \pm 0.008$ & 3.271 & r    \\\hline
      sc1   & $0.412^{+0.053}_{-0.064}$ & $0.426 \pm 0.017$ & $0.157 \pm 0.011$ & $0.128 \pm 0.005$ & $0.109 \pm 0.003$ & 3.507 & n    \\ 
      sc2   & $0.303^{+0.04}_{-0.069}$  & $0.329 \pm 0.009$ & $0.103 \pm 0.005$ & $0.153 \pm 0.016$ & $0.116 \pm 0.018$ & 1.435 & r    \\
      sc3   & $0.666^{+0.043}_{-0.033}$ & $0.34 \pm 0.022$  & $0.014 \pm 0.025$ & $0.454 \pm 0.011$ & $0.238 \pm 0.006$ & 1.789 & PS08 \\\hline
      sr1   & $0.115^{+0.023}_{-0.025}$ & $0.286 \pm 0.012$ & $0.1 \pm 0.004$   & $0.215 \pm 0.021$ & $0.228 \pm 0.017$ & 7.903 & r    \\ 
      sr2   & $0.111^{+0.035}_{-0.042}$ & $0.265 \pm 0.008$ & $0.108 \pm 0.005$ & $0.144 \pm 0.014$ & $0.162 \pm 0.019$ & 1.97  & r    \\
      sr3   & $0.149^{+0.032}_{-0.062}$ & $0.28 \pm 0.01 $  & $0.169 \pm 0.008$ & $0.1 \pm 0.008$   & $0.091 \pm 0.005$ & 1.742 & n    \\
      sr4   & $0.156^{+0.085}_{-0.077}$ & $0.406 \pm 0.009$ & $0.278 \pm 0.006$ & $0.098 \pm 0.009$ & $0.084 \pm 0.007$ & 1.323 & n    \\\hline
    \end{tabular}                                                                                                        
    \begin{tabular}{cccccc}
      \hline
      Sample & $\langle \gamma \rangle$ & $\sigma_{\gamma}$ & $\langle \epsilon \rangle$ & $\sigma_{\epsilon}$ &  $\chi^2/\textmd{d.o.f.}$\\\hline
      total & $0.584 \pm 0.006$	& $0.164 \pm 0.005$ & $0.117 \pm 0.01$  & $0.101 \pm 0.006$ & 7.998 \\\hline
      em1   & $0.543 \pm 0.007$ & $0.166 \pm 0.006$ & $0.122 \pm 0.01$  & $0.103 \pm 0.007$ & 3.358 \\
      em2   & $0.632 \pm 0.008$ & $0.155 \pm 0.007$ & $0.116 \pm 0.009$ & $0.105 \pm 0.006$ & 6.141 \\
      em3   & $0.659 \pm 0.013$ & $0.15 \pm 0.011$  & $0.11 \pm 0.009$  & $0.098 \pm 0.006$ & 4.736 \\
      em4   & $0.655 \pm 0.015$ & $0.148 \pm 0.017$ & $0.105 \pm 0.008$ & $0.093 \pm 0.007$ & 6.421 \\\hline
      ec1   & $0.59 \pm 0.004$  & $0.166 \pm 0.005$ & $0.121 \pm 0.01$  & $0.106 \pm 0.007$ & 1.645 \\ 
      ec2   & $0.579 \pm 0.006$ & $0.156 \pm 0.005$ & $0.113 \pm 0.009$ & $0.092 \pm 0.007$ & 2.489 \\
      ec3   & $0.557 \pm 0.007$ & $0.173 \pm 0.006$ & $0.108 \pm 0.009$ & $0.093 \pm 0.006$ & 4.154 \\
      ec4   & $0.577 \pm 0.007$ & $0.172 \pm 0.006$ & $0.118 \pm 0.009$ & $0.105 \pm 0.006$ & 2.619 \\\hline
      er1   & $0.525 \pm 0.007$ & $0.156 \pm 0.006$ & $0.11 \pm 0.009$  & $0.098 \pm 0.006$ & 4.437 \\ 
      er2   & $0.638 \pm 0.011$ & $0.158 \pm 0.012$ & $0.11 \pm 0.01$   & $0.093 \pm 0.007$ & 6.516 \\
      er3   & $0.675 \pm 0.019$ & $0.13 \pm 0.014$  & $0.115 \pm 0.012$ & $0.101 \pm 0.008$ & 5.625 \\
      er4   & $0.674 \pm 0.014$ & $0.151 \pm 0.013$ & $0.13 \pm 0.009$  & $0.113 \pm 0.006$ & 3.595 \\\hline
    \end{tabular}
  \end{center}
\end{table*}

\label{lastpage}

\end{document}